\documentclass{article}

\usepackage{arxiv}

\usepackage[utf8]{inputenc} 
\usepackage[T1]{fontenc}    
\usepackage{hyperref}       
\usepackage{url}            
\usepackage{booktabs}       
\usepackage{amsfonts}       
\usepackage{nicefrac}       
\usepackage{microtype}      
\usepackage{graphicx}       
\usepackage{subcaption}
\usepackage{float}
\usepackage{amsmath}        
\usepackage[numbers]{natbib}         

\usepackage{xcolor}

\DeclareMathOperator{\sech}{sech}

\title{Timescale separation enables Deep Reinforcement Learning Control of Rotating Detonation Engine Mode Transitions}
\author{
Kristian Holme\\
University of Oslo\\
holme.kristian@gmail.com
\And
Jean Rabault\\
Independent Researcher, Oslo\\
\And
Ricardo Vinuesa\\
University of Michigan
\And
Mikael Mortensen\\
University of Oslo
}

\begin{document}

\maketitle

\begin{abstract}
  Rotating detonation engines (RDEs) are a promising propulsion concept that
  may offer higher thermodynamic efficiency and specific impulse than
  conventional
  systems, but nonlinear phenomena, including transitions to
  oscillatory or chaotic
  propagation modes, can hinder practical operation.
  Deep Reinforcement Learning (DRL) has emerged as a promising method
  for controlling complex nonlinear dynamics such as those observed
  in RDEs. However, the multi-timescale nature
  of the RDE system makes direct application of DRL challenging.
  We address this challenge by reformulating the DRL problem in a
  moving reference frame that
  follows the detonation-wave pattern, making the wave structure
  appear quasi-steady to the agent.
  This reformulation enables scale separation between fast detonation
  propagation and
  slower operating-mode dynamics. We train DRL
  controllers to modulate spatially segmented injection pressure in a
  one-dimensional reduced-order RDE model
  and induce rapid transitions between different mode-locked states.
  Across a range of actuation periods, initial states, and target
  modes, controllers trained in
  the moving frame learn more reliably than those trained in a
  stationary frame and remain effective over a broader range of
  actuation periods.
  These results suggest that symmetry-aware moving reference frame
  formulations may be useful
  for related multiscale flow-control problems and that scale
  separation should be exploited
  whenever possible to enable DRL control of multi-timescale systems.

\end{abstract}

\section{Introduction}

Rotating detonation engines (RDEs) are combustion engines that utilize
constant-volume Humphrey-based cycles \citep{yunAnalysisDevelopmentTrends2024a}
to achieve higher theoretical
efficiency than traditional rocket engines, which are based on constant-pressure cycles. In theory, RDEs could achieve up to 23 \% \citep{wolańskiDetonativePropulsion2013a} higher fuel efficiency and up to 14 \% \citep{heisterRotatingDetonationCombustion2022} higher specific impulse compared to traditional rocket engines. These gains hold promise for enhancing overall propulsion efficiency in rocket and air-breathing applications, such as high-speed vehicles \citep{braunAirbreathingRotatingDetonation2013} and next-generation launch architectures.
RDEs exist in many forms and configurations, but they all operate on the same core principle: one or more detonation waves travel circumferentially around a cylindrical or annular chamber, generating high pressure and thrust, while fuel is injected radially.
A pre-detonator often initiates the detonation pattern, usually a linear detonation chamber
attached tangentially to the main detonation chamber
\citep{baiStudyInitiationCharacteristics2024}.
In practice, theoretical efficiency gains are not easy
to achieve, due to nonideal dynamics such as variations in fuel mixture,
parasitic deflagration (combustion not associated with primary 
detonation waves) and the presence of secondary parasitic waves that compete with the main detonation waves for fuel \citep{ramanNonidealitiesRotatingDetonation2023}.
Although not yet common, organizations have successfully flown RDEs on space
missions \citep{gotoSpaceFlightDemonstration2023},
engineers have incorporated them into jet engines
\citep{yunAnalysisDevelopmentTrends2024a}, and used them for power generation
through incorporation into gas
turbines \citep{wolanskiDevelopmentGasturbineDetonation2018, sousaThermodynamicAnalysisGas2017a}.

Controlling the inherent instabilities and nonideal dynamics in the RDE system is important for both 
improving theoretical understanding and enabling practical application of RDEs.
Several studies have investigated different methods for controlling RDE operation.
Previous studies have used the equivalence ratio (the ratio of the fuel-to-oxygen ratio to the stoichiometric ratio) 
to achieve transitions between different operating modes, e.g., from two co-rotating detonation waves to a 
single detonation wave \citep{dengFeasibilityModeControl2018}, or to enforce a preferred propagation direction
by controlling the geometry of the combustion chamber \citep{yangExperimentalStudyMode2024}.
Other works used the pre-detonator to continuously inject angular momentum into the main combustion 
chamber~\citep{shengActiveDirectionControl2022}, or modulated the propellant injection
pressure~\citep{rongInvestigationCounterrotatingShock2022}.

RDEs are characterized by the existence of complex detonation patterns and nonlinear dynamics, which cause instabilities and make their development challenging. Recent studies have shown deep reinforcement learning (DRL) to be an effective method to control a wide variety of nonlinear systems in fluid mechanics \citep{garnierReviewDeepReinforcement2021,vignonRecentAdvancesApplying2023,rabaultDeepReinforcementLearning2023}. It is therefore interesting to investigate whether we can use DRL as a controller to mitigate some of the challenges and instabilities present in RDEs. DRL is a data-driven decision-making framework based on trial-and-error learning and, hence, can be effective at controlling systems that are too complex and nonlinear for traditional linearization-based control algorithms and for which simpler methods such as PID control are ineffective.

More specifically, DRL has produced closed-loop controllers that outperform classical methods in several complex flow-control problems, including, e.g., the wake behind a variety of bodies in both 2D and 3D simulations covering linear as well as turbulent conditions \citep{rabaultArtificialNeuralNetworks2019,tangRobustActiveFlow2020a,renApplyingDeepReinforcement2021,chenDeepReinforcementLearningbased2023,yanStabilizingSquareCylinder2023,yanActiveFlowControl2024,suarezActiveFlowControl2025}, boundary layer and channel flows \citep{guastoniDeepReinforcementLearning2023,sonodaReinforcementLearningControl2023,zhouReinforcementlearningbasedControlTurbulent2025}, turbulent separation bubbles \citep{fontDeepReinforcementLearning2025}, Rayleigh-Bénard convection \citep{beintemaControllingRayleighBénard2020,jeonInductiveBiaseddeepReinforcement2025,vignonEffectiveControlTwodimensional2023,vasanthMultiagentReinforcementLearning2025}, drone control and navigation \citep{huangSymmetryInformedReinforcementLearning2024,tontiNavigationSimplifiedUrban2025}, rocket engines control \citep{waxenegger-wilfingReinforcementLearningApproach2021}, civil engineering \citep{yanDeepReinforcementLearningbased2025a,guoIntelligentControlStructural2025}, wing aerodynamics control \citep{vinuesaFlowControlWings2022,wangDeepReinforcementLearning2022,rennMachineLearningFlowinformed2022,garciaDeepreinforcementlearningbasedSeparationControl2025a}, and researchers continuously report more uses of DRL in the literature and are now too many to list exhaustively.

In addition to the development of control laws per se, follow-up works have focused on making these findings applicable to real-world systems by optimizing the number and placement of sensors~\citep{parisRobustFlowControl2021}, and making it possible to use states consisting of a reduced number of measurements \citep{wangDynamicFeaturebasedDeep2024}. Therefore, it appears that researchers often apply DRL to a given fluid mechanics control problem in successive studies, typically first a proof-of-concept establishing that control can be attained at all (sometimes, researchers do this assuming unrealistic state or control authority in the first early proof-of-concept studies), before follow-up works refine these studies and distill them into realistic controllers that match real-world sensing and actuation constraints. As a consequence, researchers are now increasingly applying DRL to real-world experiments \citep{fanReinforcementLearningBluff2020,fangExperimentalDeepReinforcement2026}.

Interestingly, several key factors have enabled the application of DRL to increasingly complex problems. From a DRL point of view, two main innovations have, so far, been key to enabling the use of DRL for complex flow control problems: i) the effective use and scaling of parallel environments \citep{rabaultAcceleratingDeepReinforcement2019}, and ii) the use of multi-agent reinforcement learning to enable distributed input distributed output (DIDO) control \citep{belusExploitingLocalityTranslational2019,vignonEffectiveControlTwodimensional2023}. From a CFD simulations viewpoint, drastic speedups relying on, e.g., GPU solvers have been key to enable some of the more advanced studies presented in the literature \citep{suarezActiveFlowControl2025,fontDeepReinforcementLearning2025}, which would otherwise have been computationally unfeasible. Finally, in real-world physical experiments, the move towards ultra-fast field-programmable gate array (FPGA, \citep{brownFPGACPLDArchitectures1996a}) controllers \citep{chenFieldProgrammableGate2024,zongClosedloopSupersonicFlow2025} and advanced techniques such as plasma actuators \citep{kongDeepReinforcementLearning2026,fangExperimentalDeepReinforcement2026}, is making it possible to run kHz-update-rate closed-loop control in increasingly realistic aerodynamics configurations. This large-scale, distributed, community-driven exploration of DRL control applications, ranging from idealized (and often unrealistic) proof-of-concepts in CFD simulations, to algorithmic improvements, to actual engineering and deployment in real-world experiments, has proven key to enable fast progress in the field of DRL for flow control.

In this paper, we use deep reinforcement learning to train neural
networks to actively control the injection pressure in a one-dimensional RDE
reduced-order model (ROM). To the best of our knowledge, this is the first time anyone has used DRL for this kind of RDE control application. Therefore, we focus on establishing if, and how, DRL can be effectively applied to RDE control.
Our study utilizes an idealized RDE ROM to minimize computational costs, while providing the controller with extensive sensing and control authority.
Following this proof-of-concept, future work can adapt our findings to real-world engineering constraints.
This trajectory aligns with similar advancements in other fluid mechanics control tasks, as discussed above.
In particular, we believe that the scale separation technique that we introduce in the following, while being "just" a simple change of reference frame at its core, is a key enabling idea to apply DRL to RDE control, similar to how the invariant (now called MARL) technique introduced by \citet{belusExploitingLocalityTranslational2019} has been a key ingredient to tame the curse of dimensionality that otherwise arises in DIDO problems.

More specifically, the goal of the DRL controller in our present study is
to achieve rapid transitions between mode-locked states while avoiding
chaotic wave propagation.
A mode-locked state is a stable traveling-wave solution of the RDE ROM equations in which $n$ detonation waves propagate at equal and
 constant speed with fixed angular spacing of $2\pi/n$, constituting a fixed point of the dynamical system.
A major difficulty in realizing effective DRL-based control is that the RDE exhibits several relevant timescales, which extend over multiple orders of magnitude \citep{kochMultiscalePhysicsRotating2021}.
Choosing an actuation frequency for the learned controller thus represents a tradeoff between
influencing the fast local processes, e.g., detonation wave propagation, and the slower global processes, e.g., modulations and
mode transitions of the system.
In our configuration, the timescale on which the wave pattern changes substantially can be roughly $25$ times slower than the period of a full rotation. This separation of timescales complicates learning, in particular temporal credit assignment and the choice of actuation frequency.
Here, we mitigate this challenge by exploiting the rotational symmetry of the problem to define a moving reference frame that follows the traveling detonation waves.
This reformulation effectively closes the gap between the fast propagation timescale of the rotating waves and the slower timescales associated with modulation and mode transitions.
This allows the controller to act on a slower timescale while retaining effective control over individual detonation waves that live on a fast timescale.

We evaluate the learned controllers in two complementary studies.
First, we compare a single-agent uniform (SU) controller with two-step and PID controllers under matched uniform actuation and a common evaluation protocol.
Second, we compare four DRL configurations using a segmented control strategy (SSS, SSM, MSS, MSM; \autoref{tab:naming-conventions}), where the configurations differ in reference frame and in single- versus multi-agent structure.
\begin{table}[htbp]
  \centering
  \caption{Learned controller configurations.
  The reference-frame distinction applies to segmented observations and actions.
  The SU observation is invariant to rotational phase, so a reference-frame label is not applicable.
  The letter scheme is detailed in Section~\ref{sec:agent-configs}.}
  \label{tab:naming-conventions}
  \begin{tabular}{llll}
  \toprule
  Name & Agent type & Injection pressure & Reference frame \\
  \midrule
  SU    & Single & Uniform   & N/A (rotation invariant) \\
  SSS   & Single & Segmented & Stationary  \\
  SSM   & Single & Segmented & Moving \\
  MSS   & Multi & Segmented   & Stationary  \\
  MSM   & Multi & Segmented   & Moving \\
  \bottomrule
  \end{tabular}
\end{table}

Section~\ref{sec:methodology-rde} describes the RDE model, and Sections~\ref{sec:drl-framework}--\ref{sec:experimental-setup} define the learned controllers, baselines, training setups, and evaluation protocols.
Section~\ref{sec:results} presents the uniform benchmark followed by the segmented-controller study, and Section~\ref{sec:conclusion} summarizes the conclusions.

\section{Methodology}
\label{sec:methodology}

\subsection{RDE model}
\label{sec:methodology-rde}
\subsubsection{1D model equations}

To model the RDE dynamics, we use the set of one-dimensional model equations
from \citet{kochMultiscalePhysicsRotating2021}.
The equations model the interplay between two variables, $u$ and $\lambda$,
on a periodic domain $x \in [0, L)$ with $L = 2\pi$, representing the
circumference of the annular combustion chamber.
$u$ is a property analogous to specific internal energy \citep{kochMultiscalePhysicsRotating2021} and $\lambda$ represents the
progression of combustion,
i.e., how much fuel the combustion process has consumed in a given spot.

The governing equations are:

\begin{equation}
  \begin{aligned}
  &\frac{\partial u}{\partial t} + u\frac{\partial u}{\partial x} =
  (1-\lambda)\omega(u)q_0 + \nu_1 \frac{\partial^2 u}{\partial x^2} +
  \epsilon \xi (u, u_0), \\
  &\frac{\partial \lambda}{\partial t} = (1-\lambda)\omega(u) - \beta
  (u, u_p, s)\lambda + \nu_{2}\frac{\partial^2 \lambda}{\partial x^2},
  \\
  &\omega(u) = e^{(u-u_c)/\alpha},\,
  \beta (u, u_p, s) = \frac{su_p}{1+e^{r(u-u_p)}},\,\text{and}\,
  \xi (u, u_0) = (u_0-u)u^n,
  \end{aligned}
  \label{eq:rde_equations}
\end{equation}

\noindent where $\omega$ is the combustion rate; $u_c$ is the critical threshold above which significant reaction occurs,
and $\alpha$ controls the width of the reaction zone.
$q_0$ is the heat release of the propellant. The refueling rate $\beta$ depends 
on $u_p$ (injection pressure), $s$ (refueling strength), 
and $r$ (a steepness parameter for pressure-coupling sensitivity: a higher $r$ sharpens the transition between full refueling when $u \ll u_p$ and quenching when $u \gg u_p$).
The term $\xi$ models dissipation, with $u_0$ the ambient reference value 
and $\epsilon$ a loss coefficient.
Following~\citet{kochMultiscalePhysicsRotating2021}, we include diffusion terms, with coefficients $\nu_1$ and $\nu_2$ for $u$ and $\lambda$, respectively. $u_p$ is the fuel injection pressure. In the original formulation of \citet{kochMultiscalePhysicsRotating2021}, $u_p$ is a uniform scalar. In the present work, we generalize it to a spatially and temporally varying function $u_p(x, t)$, enabling segmented control of injection pressure across the domain. We restrict $u_p(x,t) \in [0, 1.2]$ everywhere.

We adopt the default parameters from
\citet{kochMultiscalePhysicsRotating2021}, shown in \autoref{tab:rde_params}.

\begin{table}[tbp]
    \centering
    \caption{Parameter values used in the RDE model equations.}
    \renewcommand{\arraystretch}{1.2}
    \begin{tabular}{ccccc}
      \toprule
      $u_c$ & $\alpha$ & $q_0$ & $u_0$ & $n$ \\
      \midrule
      1.1 & 0.3 & 1.0 & 0.0 & 1 \\
      \addlinespace
      $\nu_1$ & $\nu_2$ & $\epsilon$ & $r$ & $s$ \\
      \midrule
      0.0075 & 0.0075 & 0.15 & 5.0 & 3.5 \\
      \bottomrule
    \end{tabular}
    \label{tab:rde_params}
\end{table}

\subsubsection{Numerical solver}

We solve the model equations \eqref{eq:rde_equations} on a uniform grid of $N = 512$ cells with periodic boundary conditions using a conservative finite-volume (FV) method\footnote{The code is available at
  \href{https://github.com/KristianHolme/RDE.jl}{github.com/KristianHolme/RDE.jl}}.
We discretize the advection term $u\partial u / \partial x$, corresponding to the flux $f(u) = u^2/2$, using MUSCL (Monotonic Upstream-centered Scheme for Conservation Laws) reconstruction with a monotonized central (MC) slope limiter \citep{vanLeerUltimateConservativeDifference1977}, and a Rusanov (local Lax--Friedrichs) numerical flux for the interface values. We discretize diffusion terms with standard second-order central differences. We perform time integration using the third-order strong stability preserving Runge--Kutta method (SSPRK33) \citep{shuEfficientImplementationEssentially1988} through the \texttt{OrdinaryDiffEq.jl} package \citep{rackauckasDifferentialEquationsjlPerformantFeatureRich2017} in the Julia programming language~\citep{bezansonJuliaFreshApproach2017}. A Courant-Friedrichs-Lewy (CFL) condition that accounts for advection, diffusion, and reaction timescales limits the time step, using a safety factor of $0.62$.

We use this combination as it is a standard conservative discretization for problems with shock-like structures: the limited MUSCL reconstruction reduces spurious oscillations near steep fronts, while the Rusanov flux provides a simple stabilizing numerical dissipation. SSPRK33 time integration is a common pairing that preserves strong stability properties under the CFL-limited explicit time step.

\subsubsection{RDE system dynamics}
\label{sec:system-dynamics}

For a detailed analysis of the behavior of the RDE dynamics, 
we refer the reader to \citet{kochMultiscalePhysicsRotating2021}.
Here, we mention the most important aspects for our present work.

The detonation waves inside an RDE exhibit a variety of behaviors.
Although the nonlinear dynamics underlying RDE operation are not
completely understood, several studies have investigated the
characteristic operational behaviors of RDEs.
A main characteristic is the so-called fundamental instability
\citep{kochMultiscalePhysicsRotating2021}.
The instability is a Hopf bifurcation, where a fixed point of the dynamical system loses stability and evolves into a periodic orbit.
For the RDE system, this entails moving from a mode-locked quasi-steady state,
to a state of time-periodic modulation of wave amplitude, wave speed, and the phase difference between waves.
This qualitative change in operating mode can be the result of a change in injection pressure.

As the injection pressure is modified further, the system may break out of
this oscillatory state and move into a new mode-locked state.
Similar dynamics occur in laser cavities, where the cavity amplifies a single pulse and then splits it into two or more pulses
\citep{liGeometricalDescriptionOnset2010, kochModeLockedRotatingDetonation2020}.
\citet{kochMultiscalePhysicsRotating2021} establish that a balance between heat release
(\emph{gain}), exhaust and expansion processes (\emph{dissipation}), and propellant injection and mixing (\emph{gain recovery})
determines the global gain dynamics that govern the RDE operation.
In this context, \emph{gain} refers to the energy added to the chamber by heat release when newly mixed propellant burns.
Local reactant availability limits instantaneous gain; combustion (both detonative burning and parasitic deflagration) depletes this availability, and injection and mixing replenish it on slower timescales.
Sources of dissipation include exhaust and expansion out of the chamber and heat transfer to the walls of the combustion chamber.

A key element in the dynamics is how the detonation waves affect the
injection of propellant.
The design of fuel and oxidizer injectors differs, but they are
typically choked
orifices~\citep{luRotatingDetonationWave2011}.
Despite this, the high pressure of the detonation waves can
locally inhibit the flow of propellants, or even cause backflow.
This coupling between heat release and propellant injection/mixing (gain recovery), mediated by injector blocking and refill, is central to the RDE dynamics.
This is the reason for calling the instability described above a fundamental instability.
In other words, some of the unwanted nonidealities of the RDE system are intrinsic to the core dynamical system.

The elements of the gain dynamics also give rise to different timescales.
\citet{kochMultiscalePhysicsRotating2021} establish four timescales, ranging from the fastest timescale of combustion to the several orders of magnitude slower timescale of dissipation.
The following timescales are expressed in the nondimensional time units of the model equations \eqref{eq:rde_equations}:
\begin{enumerate}
    \item \textbf{Combustion timescale} (fastest): governed by the reaction rate $\omega(u)$; a detonation front passes a given point in $\approx 10^{-2}$ time units.
    \item \textbf{Wave propagation timescale}: a single detonation wave traverses the domain ($L = 2\pi$) in approximately $2\pi / D \approx 3.7$ time units, where $D$ is the wave speed.
    \item \textbf{Gain recovery timescale}: refueling of the consumed propellant, governed by $\beta$; on the order of $1$--$10$ time units, comparable to the wave propagation timescale.
    \item \textbf{Dissipation timescale} (slowest): exhaust/expansion and other loss processes; participates in the slow gain--loss balance that governs instabilities and mode transitions, on the order of $10$--$10^2$ time units.
\end{enumerate}

Crucially, the dynamics of mode transitions usually live on the slowest timescale, while the structures to be controlled, i.e., the detonation waves, live on the intermediate and fast timescales (wave propagation and combustion timescales). This timescale separation is a central challenge for learning effective control strategies and motivates the moving reference frame approach introduced in Section~\ref{sec:moving-frame}.

\begin{figure}[tbp]
  \centering
  \includegraphics[width=16cm, height=12cm]{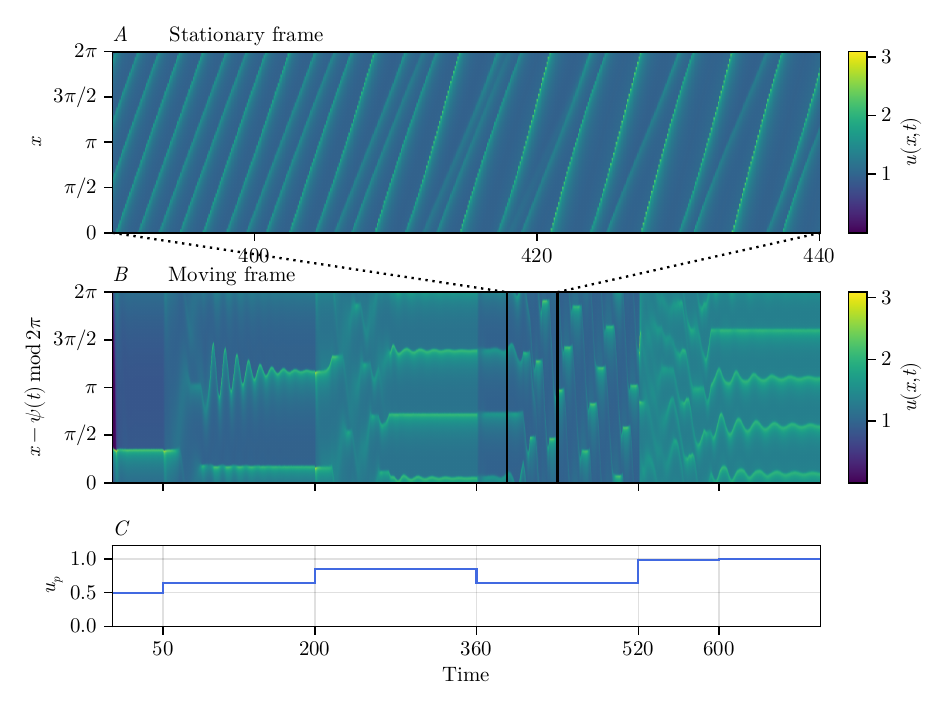}
  \caption{Simulation of \eqref{eq:rde_equations} with a predefined
    step-controller.
    Starting from an initial bump in $u$, defined by $u(x) =
    \frac{3}{2} \sech^{20}(x-1)$, and spatially uniform combustion progress $\lambda(x) = 0.5$,
    the system transitions to a single detonation wave stably
    propagating. As we increase $u_p$, the system bifurcates into a
    state of two detonation waves with oscillatory phase difference.
    Before the phase difference stabilizes at a value of $\pi$, we increase the pressure again, at $t=200$. At this point, we see a
    similar bifurcation to three waves, and the phase differences
    between them almost stabilize at a value of $2\pi/3$. As
    we decrease $u_p$ at $t=360$ the system enters a time-periodic
    oscillatory state with two galloping detonation waves. Here, the
    stronger wave is moving faster than the weaker wave, traveling
    around the domain until it catches up to the weaker wave, and
    wave strength is exchanged. The moving reference frame follows
    the fastest and strongest
    wave. From this point of view, we see that the weaker wave is
    traveling backwards across the domain until it is in front of the
    stronger wave. At this point the strengths of the waves shift, and
    the previously stronger wave weakens and repeats the pattern. After a
    series of increases of $u_p$ from this state, the system reaches a
    mode-locked state with four stably propagating waves.
    (a): $u(x,t)$ shown in the stationary reference frame where $t \in [380, 440]$; the rapid movement of the detonation waves across the domain is visible.
    (b): $u(x,t)$ in a moving reference
    frame with position $\psi$ (see Section~\ref{sec:moving-frame} for details on the moving reference frame).
    (c): Injection pressure $u_p(x,t)$ as a function of the non-dimensional time units.}
  \label{fig:sample-run}
  \phantomsubcaption{}\label{fig:sample-run-a}
  \phantomsubcaption{}\label{fig:sample-run-b}
  \phantomsubcaption{}\label{fig:sample-run-c}
\end{figure}

To illustrate these phenomena in a more specific way, we can visualize some key patterns emerging from
\eqref{eq:rde_equations}. Starting from a wide range of initial conditions of $u$
and $\lambda$, and a uniform injection pressure of $u_p = 0.5$, the system
stabilizes after some time (depending on the initial condition)
in a mode-locked state with a single traveling detonation wave.
From this state, steadily increasing $u_p$ leads to an increase
in the wave speed and amplitude. Eventually, the system reaches a
bifurcation point, where the single detonation wave breaks down and two
detonation waves emerge. Between the mode-locked states of one and two
detonation waves, we may reach an oscillatory limit cycle state of galloping
waves -- two waves,
one strong and one weak, taking turns chasing each other and
exchanging strength.
This galloping behavior can be seen in \autoref{fig:sample-run}
around $t=400$. Note that the plot in \autoref{fig:sample-run-b} is in a reference frame moving together with the detonation waves, in contrast to the excerpt in \autoref{fig:sample-run-a}, which we visualize in a stationary reference frame. This visualization in a moving reference frame is usually easier to interpret and clearer, and therefore we use it in most of the figures in the present work. We go into more detail on the moving reference frame in Section~\ref{sec:moving-frame}.

Multi-stability and hysteresis are another important aspect of the RDE dynamics, also visible in \autoref{fig:sample-run}. When $t \in [50, 200]$, we set the injection pressure to $u_p=0.64$, and we observe that the system converges to a mode-locked state towards the end of the time interval.
We apply the same injection pressure in the interval $t \in [360, 520]$, but here the system does not reach the same mode-locked state. Instead, the system enters the galloping state as previously described.

\begin{figure}[tbp]
  \centering
  \includegraphics[width=16cm, height=8cm]{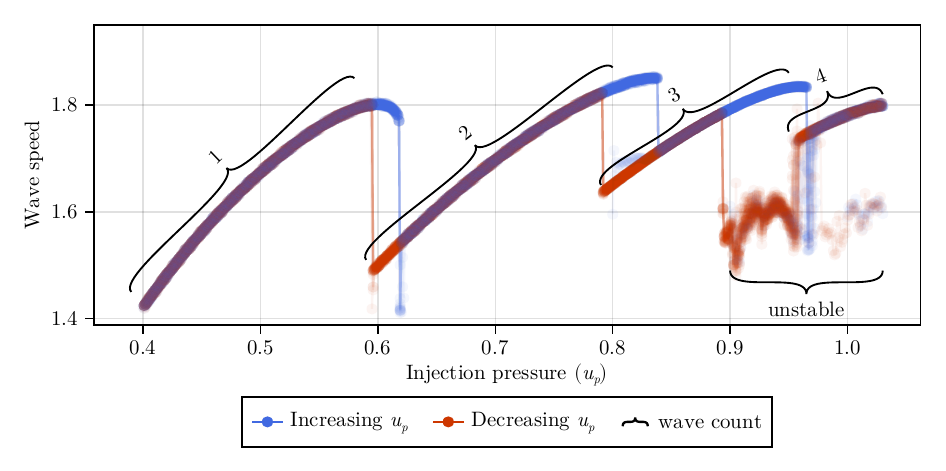}
  \caption{Wave speed as a function of injection pressure $u_p$, during upwards and downwards sweeps of $u_p$.
    The response is qualitatively similar to Figure~12 in \citet{bykovskiiContinuousSpinDetonation2006}.
    We begin each simulation by initializing $u(x,t)$ with a localized bump.
    The injection pressure $u_p$ is then uniformly stepped up or down to one of thirteen evenly spaced setpoint values in the interval $[0.4, 1.03]$.
    From each setpoint value, we sweep $u_p$ continuously through a prescribed cycle; up to $1.03$, down to $0.4$, and back to the setpoint value. The injection pressure is kept steady for $300$ time units after each step, to allow the system to stabilize. At the end of each stabilization phase, we record the average wave speed and the injection pressure as a single data point.
    In overlap regions where multiple regimes coexist, the system exhibits hysteresis: the wave count depends on the direction of the pressure sweep.
    When $u_p$ is decreased from a four-wave state, the solution passes through a band of non-mode-locked behavior before settling into a stable three-wave state.
    In this transitional band, waves form and extinguish rapidly, making speed detection unreliable. A small number of affected data points have been removed as outliers.
    Simulations were run with $N=2048$ grid points to maximize the accuracy of speed detection.}
\label{fig:wave-speed}
\end{figure}

The multi-stability of the RDE system is also seen in \autoref{fig:wave-speed}, which
plots wave speed against injection pressure $u_p$ as $u_p$ is slowly ramped up and down.
Along each branch, increasing $u_p$ raises the wave speed until a bifurcation
is crossed and the wave count increases and the wave speed decreases.
Different branches correspond to mode-locked states with different
numbers of detonation waves; multi-stability appears where these branches overlap
in $u_p$. Generally symmetric behavior is observed when decreasing $u_p$, though complex hysteresis is also observed, and the system becomes especially complex and unstable for high values of $u_p$.

The hysteresis exhibited by the RDE system is also visible in \autoref{fig:wave-speed}.
Building on the wave-count branches described above, several injection-pressure
intervals exhibit overlapping responses with different numbers of detonation
waves. The system traverses some segments only when ramping $u_p$ up versus down, producing hysteresis loops.
This hysteresis and unstable dynamics complicate the search for and design of control strategies
\citep{kochModeLockedRotatingDetonation2020,
  kochMultiscalePhysicsRotating2021,
ramanNonidealitiesRotatingDetonation2023}. When
adjusting the injection pressure to transition to one mode-locked
state (e.g., four detonation waves), the control strategy must be
adjusted depending on the starting state. Rapidly increasing the
injection pressure from a state of three detonation waves to a value
capable of sustaining four detonation waves does not necessarily
yield the same result if starting from a state of two detonation waves.
Similarly, even though an injection pressure value of $1.02$ could sustain
four detonation waves, adjusting the pressure directly to this value
may not result in four stable detonations but may lead to
extinction of the initial detonation waves and result in only deflagration
or a combination of deflagration (flame) fronts and detonation waves that compete
for fuel in such a way
that none of the waves manage to become stable detonations.

These complex nonlinear dynamics are a key challenge in the development of RDEs. The combination of nonlinearity, hysteresis, and different timescales makes the development of control strategies challenging, and is the key motivation for investigating the possible use of DRL for RDE control, since DRL has, as discussed above, shown promising results in a variety of complex control tasks.

\subsection{Deep reinforcement learning}
\label{sec:drl-framework}

Deep reinforcement learning is a field at the intersection of reinforcement learning (RL) and deep learning (DL).
Reinforcement learning is an approach to learn decision-making strategies that maximize a reward signal. RL is formalized into two interacting parts, an agent and an environment. The environment passes an observation, i.e., a full or partial description of the state of the environment, to the agent. The agent then produces an action that it executes in the environment, transitioning the environment to a new state.
Based on the previous environment state, the new state, and the supplied action, the environment calculates a numeric reward $r_t$ at the time step $t$ and gives it back to the agent to estimate its performance.
The agent then uses the reward signal to adjust its decision-making process, reinforcing decisions leading to high rewards and punishing decisions that lead to low rewards.
By focusing on maximizing the reward function, RL is a goal-oriented learning approach.
The RL formulation does not need any knowledge about how to make a decision, it only needs to be able to judge the agent's decisions and interact with and control the system through trial-and-error learning.

For a more detailed presentation of reinforcement learning, we refer the reader to \citet{suttonReinforcementLearningIntroduction2018a}. Since this topic has been covered many times also in the fluid mechanics literature, as discussed above, we only provide a general overview of DRL and refer readers interested in technical and implementation details to previous works that discuss these details thoroughly.

Deep learning is a discipline where artificial neural networks are used as universal function approximators to learn a desired function.
We combine deep learning with RL by letting neural networks do the decision-making
and using gradient-based optimization methods to learn the optimal network weights to produce actions that maximize the reward function.
In this paper, we use the Proximal Policy Optimization (PPO) algorithm \citep{schulmanProximalPolicyOptimization2017}. PPO is a widely used on-policy actor-critic policy gradient method, which researchers have employed in many of the applications listed in the introduction. These three characterizing terms of PPO correspond to the following characteristics:

\textbf{Policy gradient:} The agent uses a neural network that maps the environment observation directly to an action. This network is referred to as the actor network, or simply as "the actor". 

\textbf{Actor-critic:} In addition to the actor, the agent has a value network that estimates the expected future reward for the current observation. This network is referred to as the critic network. The agent uses the critic's estimate of the future reward to judge whether the observed outcome was better or worse than expected and to update the actor accordingly.

\textbf{On-policy:} Training a DRL agent consists of repeatedly collecting data from interacting with the environment and then training the neural networks. In on-policy algorithms, the collected data is discarded after each training phase. In other words, all data used in a training phase are collected using the weights active at the beginning of that phase.

Most modern policy-gradient methods, including PPO, use a stochastic policy: at each step, the agent randomly samples an action from a learned distribution. This promotes exploration during data collection.

When calculating the value at a certain step in the agent-environment interaction, a factor $\gamma \in [0,1)$ discounts future rewards $r_{t+k}$, where $t$ is the current time step and $k\ge 1$. The discounted return, including the undiscounted current reward $r_t$, is given by
\[
G_t = \sum_{k=0}^{\infty} \gamma^k r_{t+k},
\]
\noindent so $\gamma$ controls how far into the future the agent accounts for the consequences of its control. Smaller $\gamma$ (positive close to $0$) emphasizes immediate rewards, whereas $\gamma$ close to $1$ makes the agent more farsighted. A common rule of thumb is an effective horizon of order $1/(1-\gamma)$ steps, e.g., with $\gamma = 0.99$, the agent takes approximately $1/(1-0.99)=100$ future steps into account. This horizon determines the temporal window over which the agent attempts to assign credit to its actions \citep{pignatelliSurveyTemporalCredit2023}. A longer horizon allows the agent to account for more distant consequences, but also increases the number of actions that compete for credit assignment and, therefore, makes learning more difficult \citep{pignatelliSurveyTemporalCredit2023, laidlawBridgingRLTheory2023}.

When the action period $\Delta t$ (the simulation time between successive decisions) varies between 
experimental configurations, a fixed $\gamma$ would produce different physical time horizons. 
To ensure a consistent \emph{physical reward horizon} $T_h$ within each training setup, we set
\begin{equation}
    \gamma = 1 - \frac{\Delta t}{T_h},
    \label{eq:gamma}
\end{equation}
\noindent so that the effective horizon in steps is $1/(1-\gamma) = T_h / \Delta t$, 
corresponding to a fixed physical time window of $T_h$ time units regardless of the actuation frequency.

For the segmented setup, we use $T_h = 10$, which yielded better training performance than
using a fixed $\gamma$ over action periods. This choice places the reward horizon on the 
gain recovery timescale (see Section~\ref{sec:system-dynamics}), ensuring that agents consider the slow dynamics relevant to mode transitions.
The separate SU setup uses $\Delta t=10$ and a longer physical reward horizon of $T_h=100$, giving $\gamma=0.9$, to cover the slower response available through uniform actuation.

This physical time horizon creates a fundamental challenge for agents to learn effective control strategies:
 a short $\Delta t$ requires many steps ($T_h / \Delta t$) to span the same physical horizon,
 meaning many more actions compete for credit assignment. In the temporal credit assignment literature,
  this corresponds to an increase in \emph{Markov decision process (MDP) depth},
a key source of difficulty that increases variance and makes it harder to attribute outcomes to the 
responsible actions \citep{pignatelliSurveyTemporalCredit2023,laidlawBridgingRLTheory2023}.
This tension is a key motivation for the moving reference frame approach (Section~\ref{sec:moving-frame}),
 which mitigates the credit assignment challenge by reducing the effective complexity of the control problem.

We use a custom PPO implementation for the DRL training algorithm in the Julia programming language\footnote{Code is available at
  \href{https://github.com/KristianHolme/Drill.jl}{github.com/KristianHolme/Drill.jl}
}.

\subsection{Environment design}

In this paper, we focus on using DRL to achieve rapid transitions between mode-locked states, avoiding nonideal oscillatory or galloping states and chaotic propagation.

\begin{figure}[tbp]
    \centering
    \includegraphics[width=8cm, height=10cm]{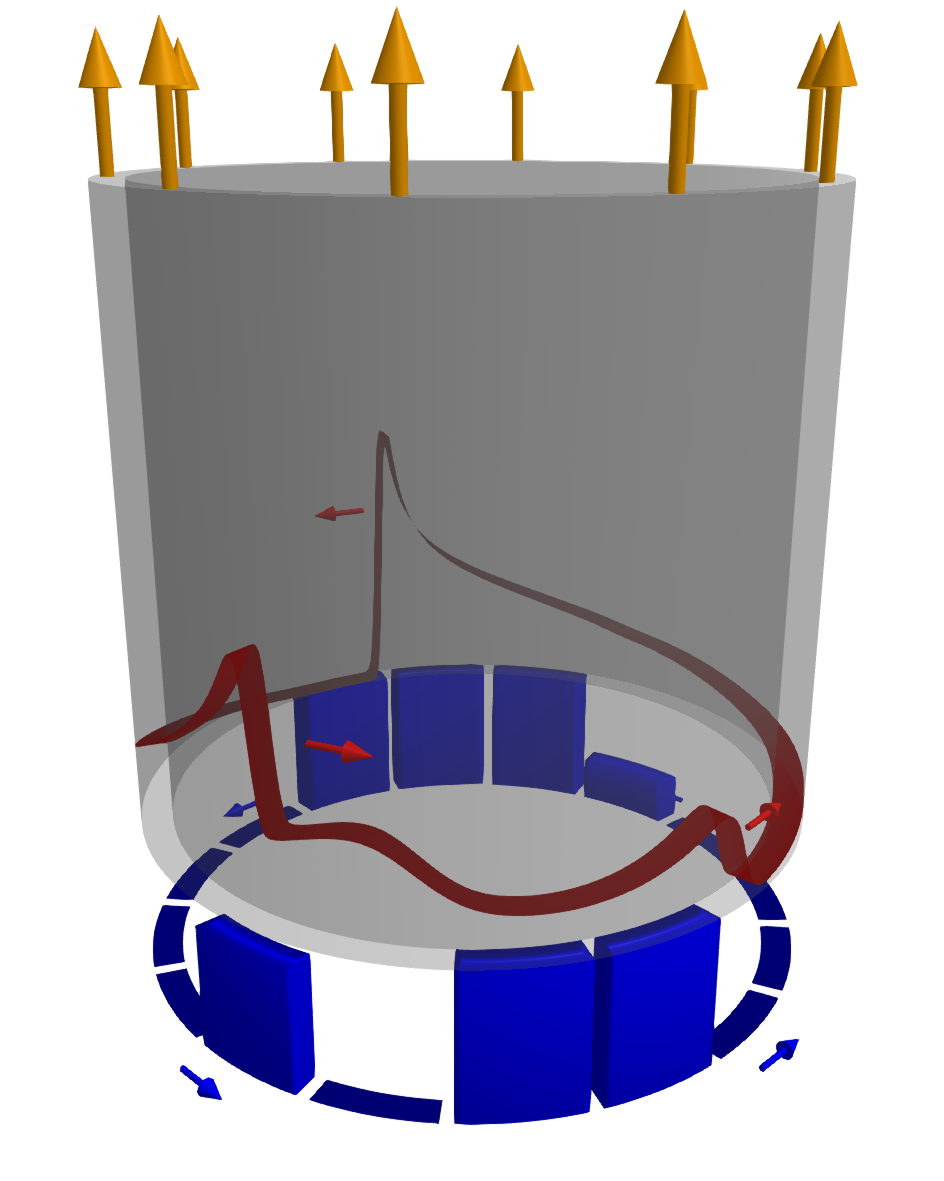}
    \caption{Illustration showing one-dimensional RDE simulation data mapped onto a three-dimensional annular cylinder.
     The red band shows the $u$ variable, with red arrows indicating counterclockwise motion.
     The blue bars illustrate sections with individually controlled injection pressure,
      with blue arrows indicating that the sections are attached to the traveling reference frame.
       Yellow arrows illustrate gas exhaust/expansion.}
    \label{fig:rde_diagram}
\end{figure}

The setup of the environment with segmented injection pressure control, where the control sections follow the moving reference frame, is illustrated in \autoref{fig:rde_diagram}.
Note that this moving reference frame approach is mainly a reformulation to exploit symmetry and separate timescales.
Implementing an injection pattern that effectively travels with the detonation structure would likely require substantial actuation bandwidth and additional engineering, and we do not claim that it is directly realizable in current RDE hardware, where injectors are fixed in the laboratory frame.
Sensing a usable mean wave speed for the frame estimate is a related but distinct challenge. We return to both sensing and actuation feasibility after describing the algorithm in Section~\ref{sec:moving-frame}.
As previously discussed and as visible in earlier DRL flow control studies, the present formulation is an effective way to evaluate the controllability of the system given maximal control authority.
We therefore present the moving-frame segmented controller as a controllability proof-of-concept under maximal control authority, not as a near-term hardware-ready design.

Deploying a learned controller requires sensing, inference, and actuation at the action rate, and these requirements differ between the uniform and segmented configurations.
Under a representative laboratory scaling (circumference $0.224\,\mathrm{m}$, wave speed $\approx 1250\,\mathrm{m\,s^{-1}}$~\citep{kochDatadrivenSurrogatesRotating2021}, revolution period $\approx 180\,\mu\mathrm{s}$), the SU action period $\Delta t = 10$ corresponds to an update rate of roughly $2\,\mathrm{kHz}$ (\autoref{tab:physical-rates}).
A single wall-mounted ion probe sampled at $250\,\mathrm{kHz}$~\citep{johnsonUseConvolutionalNeural2023} records on the order of 45 samples per revolution.
For a wave pattern that changes slowly over one revolution, the spectrum of such a signal could partially recover the rotation-invariant Fourier magnitudes used as the SU observation, although the combustion progress $\lambda$ is not directly measurable.
Inference with a small network at kilohertz rates appears within reach of FPGA-based DRL control~\citep{zongClosedloopSupersonicFlow2025}.
Actuator bandwidth is likely the more restrictive constraint, because the present simulations apply each action over a short smoothing ramp that is faster than typical valve and plenum response.
The moving-frame configurations impose a stricter actuation requirement (\autoref{tab:physical-rates}; Section~\ref{sec:moving-frame}).

\begin{table}[tbp]
  \centering
  \caption{Approximate physical update and actuation rates under a representative laboratory scaling
  (circumference $0.224\,\mathrm{m}$, wave speed $\approx 1250\,\mathrm{m\,s^{-1}}$~\citep{kochDatadrivenSurrogatesRotating2021}),
  with one simulation time unit $\approx 49\,\mu\mathrm{s}$ at the mean model speed $D = 1.71$.
  Rates are order-of-magnitude estimates.
  All configurations apply each action over a smoothing ramp of $\tau_{\mathrm{smooth}} = 0.1$ ($\approx 5\,\mu\mathrm{s}$).
  For the moving-frame controllers, the lower value corresponds to eight injection switches per revolution for the four-lobed bang--bang template in \autoref{fig:sss-ssm-3to4}, and the upper value to one update per grid-cell passage ($N = 512$).}
  \label{tab:physical-rates}
  \begin{tabular}{llll}
  \toprule
  Controller & Action period $\Delta t$ & Policy update rate & Injector modulation rate \\
             & {\footnotesize\color{gray} simulation time} & {\footnotesize\color{gray} physical} & {\footnotesize\color{gray} physical} \\
  \midrule
  SU         & $10$          & $\approx 2\,\mathrm{kHz}$          & $\approx 2\,\mathrm{kHz}$ \\
  SSS, MSS   & $0.25$--$8.0$   & $\approx 80$--$2.5\,\mathrm{kHz}$  & $\approx 80$--$2.5\,\mathrm{kHz}$ \\
  SSM, MSM   & $0.25$--$8.0$   & $\approx 80$--$2.5\,\mathrm{kHz}$  & $\approx 45\,\mathrm{kHz}$--$3\,\mathrm{MHz}$ \\
  \bottomrule
  \end{tabular}
\end{table}

In the following, we discuss the technical details of the environment setup and coupling to the DRL controllers.

\subsubsection{Observation space}
\label{sec:observation-space}

\begin{figure}[tbp]
    \centering
    \includegraphics[width=16cm]{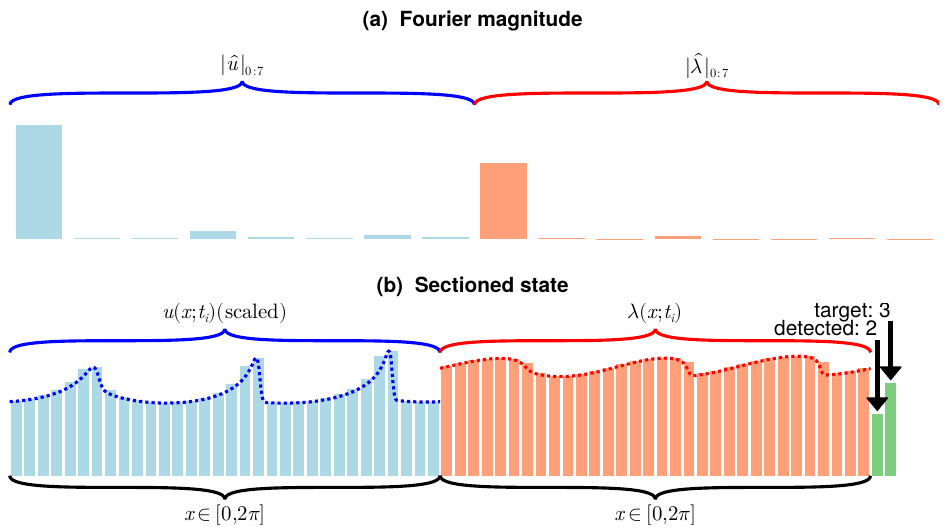}
    \caption{Observation encodings used by the learned controllers.
    (a)~The SU observation contains the magnitudes of the lowest eight real Fourier coefficients of $u$ and $\lambda$.
    The magnitudes are invariant to circular translations, and truncation removes higher spatial frequencies.
    (b)~The segmented observation contains the maxima of scaled $u$ and $\lambda$ in 32 bins, followed by the detected and target wave counts.
    Both counts are divided by four.
    The dotted curves show the fields from which the observation is constructed. In this example, the detector identifies two of the waves as detonations, and the target is three.
    Single-agent segmented controllers receive the ordered whole-domain vector.
    Multi-agent controllers receive a separate circularly recentered copy for each control segment, and moving-frame controllers shift the fields before constructing the displayed encoding.}
    \label{fig:observation-diagram}
\end{figure}

The uniform and segmented control problems use different observation encodings, as illustrated in \autoref{fig:observation-diagram}.
For the SU controller, the observation contains the magnitudes of the lowest eight real Fourier coefficients of each field, $u$ and $\lambda$, for a total of 16 entries.
A circular translation changes the phases of the Fourier coefficients but not their magnitudes.
The representation is therefore invariant to the rotational phase of the wave pattern.
Retaining only the lowest coefficients also suppresses high spatial frequencies and gives the controller a coarse representation of the global state.
Preliminary screening of uniform-controller observations found this representation to produce the most reliable training among the encodings examined.

For segmented control, we partition $[0,2\pi)$ into $M_{\mathrm{obs}}=32$ equal-width observation bins.
We record the maximum of scaled $u$ and the maximum of $\lambda$ in each bin, producing 64 field-summary entries.
We append the detected number of detonation waves\footnote{Detonation fronts are detected as large negative gradients in $u$ using a forward finite difference and an adaptive threshold, where nearby detections are merged.} and the target wave count, both divided by four.
The resulting observation has 66 entries and summarizes the full periodic domain.

For multi-agent configurations (see Section~\ref{sec:agent-configs}), we divide the environment into $M$ pseudo-environments
that share the same underlying physical simulation but each output a scalar action.
Each pseudo-environment receives the same whole-domain summary circularly recentered on its own control segment.
This centering makes the shared policy actuator-relative and removes the need to identify an absolute segment position.
Preliminary tests with a normalized segment index and sinusoidal positional encoding did not improve performance.
For moving-frame segmented controllers, we first shift the fields into the moving frame and then construct the single- or multi-agent observation.

\subsubsection{Action space and control smoothing}

The actor network of the agent outputs values in $[-1,1]$. We then linearly transform these values to correspond to $u_p$-values in $[0, 1.2]$. After each step in the environment, the agent computes a new action and we update injection pressures. To avoid discontinuities in time, we apply a smooth temporal transition between the previous and updated injection pressure profiles, $u_p^{\mathrm{prev}}$ and $u_p^{\mathrm{new}}$,  during the simulation sub-steps comprising the environment step. Specifically, the injection pressure at time $t$ during a transition is

\begin{equation}
    u_p(t) = u_p^{\mathrm{prev}} + \bigl(u_p^{\mathrm{new}} - u_p^{\mathrm{prev}}\bigr)\, g\!\left(\frac{t - t_{\mathrm{ctrl}}}{\tau_{\mathrm{smooth}}}\right),
    \label{eq:smoothing}
\end{equation}

\noindent where $t_{\mathrm{ctrl}}$ is the time of the most recent control update, and $g(x) = f(x)/\bigl(f(x) + f(1-x)\bigr)$ with $f(x) = \exp(-1/x)$ for $x > 0$ and $f(x) = 0$ otherwise, which defines a $C^{\infty}$ function that transitions smoothly from $0$ to $1$ over the interval $[0, 1]$. We use $\tau_{\mathrm{smooth}} = 0.1$ time units. For segmented control, we apply this transition independently to each control segment, i.e.,\ $u_p(x, t)$ transitions component-wise from $u_p^{\mathrm{prev}}(x)$ to $u_p^{\mathrm{new}}(x)$. In this paper, we divide the domain into $M = 16$ control segments when not using a uniform injection pressure.
This choice was informed by preliminary exploratory runs over $M \in \{4,8,16,32,64\}$, which suggested $M=16$ as a tractable compromise between spatial control authority and the dimensionality of the single-agent action space. Those sweeps are not presented as a formal ablation here.
We emphasize that the linear map gives each segment independent access to the full interval $[0,1.2]$ at every control update (subject only to the temporal blending in \eqref{eq:smoothing}).

\subsubsection{Reward function}
\label{sec:reward-function}

\begin{figure}[tbp]
    \centering
    \includegraphics[width=1\linewidth]{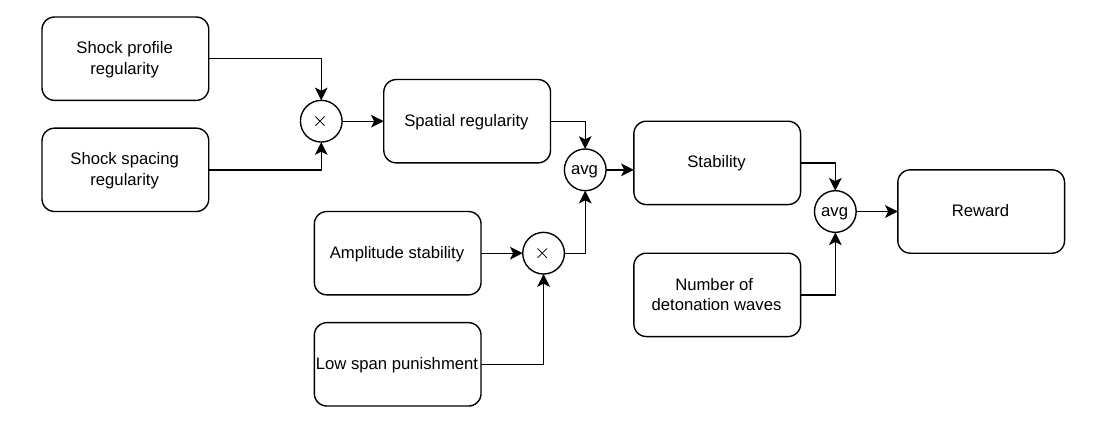}
    \caption{Reward function used during training. Each rectangle denotes a sub-reward with a range of $(0,1]$ or $[0,1]$. The reward combines a stability term, which favors temporally and spatially regular wave patterns, with a target term, which rewards matching the desired detonation-wave count. The flowchart highlights that successful controllers must both reach the target mode and keep the resulting wave pattern stable.}
    \label{fig:reward-flowchart}
\end{figure}

For the results presented here, we use the reward function summarized in \autoref{fig:reward-flowchart}. The training reward is the equally weighted sum of a \textbf{stability reward} and a \textbf{target reward}.
Where sub-rewards take values in $[0,1]$ or $(0,1]$, we multiply components that are jointly required, so that a high value of one factor cannot compensate for a low value of the other, and we average components that represent independent desirable properties.

We compute the \textbf{stability reward} over the simulation trajectory within one environment step (the full action period $\Delta t$). It is the average of an amplitude-stability component and a spatial-stability component:
\begin{itemize}
    \item \textit{Amplitude stability:} This term penalizes temporal variation in $\max(u)$ and $\min(u)$ over the environment step using an exponential decay $\exp(-\alpha_v \cdot \mathrm{R_r})$, where $\mathrm{R_r}$ is the relative range of whichever of the two extrema varies more. We multiply the resulting value by a small-span penalty $\min(1, S_p/0.1)$, where $S_p$ is the maximum span of $u$ over the time step. The threshold $0.1$ lies clearly below any shock-bearing span, so flat solutions without detonation waves are penalized.
    \item \textit{Spatial stability:} This term is the product of a periodicity reward and a detonation-front-spacing reward. The periodicity reward measures the $n_{\mathrm{target}}$-fold rotational symmetry of $u(x)$ by comparing $u(x)$ with circularly shifted copies $u(x + kL/n_{\mathrm{target}})$ for $k = 1, \dots, n_{\mathrm{target}}-1$. The detonation-front spacing reward measures the deviation of the detected front spacings from the ideal spacing $L/n_{\mathrm{target}}$. We scale and map both quantities through exponential decays, and retain the less favorable value in each case. Small floors ($\approx 0.037$ on the normalized circular-shift residual and $\approx 7.8\times 10^{-3}$ on spacing, of order the grid cell size $\mathrm{d}x = 2\pi/512$) zero out residual mismatch when the target wave count does not divide the mesh evenly, in particular for three waves.
\end{itemize}

The \textbf{target reward} is a binary indicator: $1$ if the current detonation-wave count (estimated from detonation-front detection) is equal to the target $n_{\mathrm{target}}$, and $0$ otherwise.

This composite form was refined iteratively by inspecting trajectories under fixed and interactive control and adjusting the signal until it tracked mode quality and stability clearly without excessive noise.
Sparse success-style signals based mainly on shock count or delayed transition detection proved too uninformative for training.
Simpler variants were also susceptible to reward hacking.
Some policies appeared regular only at action endpoints while evolving irregularly during the action interval, and flat high-pressure solutions scored well on amplitude and symmetry-like terms.
Short end-of-period stability windows ignored instability earlier in long actions, and incorrect wave counts were insufficiently penalized.
Judging stability over the full action period, the small-span penalty, and the equal-weight binary target term address these failure modes.
The same reward family is used in both training setups.
Training uses $\alpha_v=4$ together with the target term, while the transition evaluation defined in Section~\ref{sec:transition-evaluation} uses the stability component with $\alpha_v=1$.

\subsubsection{Moving reference frame}
\label{sec:moving-frame}

Initial segmented-control experiments showed a conflict between the update rates needed for fast wave tracking and slow mode-transition dynamics.
Frequent control updates resolve wave motion but deepen the temporal credit-assignment problem.
Sparser control updates shorten the decision sequence but cannot maintain a wave-relative segmented action in stationary coordinates.
We address this conflict by expressing segmented observations and actions in a reference frame that follows the mean detonation-wave motion.
The transformation makes the propagating wave pattern appear approximately stationary to the policy.

Constructing such a frame requires an estimate of the mean detonation wave speed $D$.
Related wave-alignment methods exist for RDE data, notably the UnTWIST algorithm of Mendible et al.~\citep{mendibleDimensionalityReductionReducedorder2020}, which shifts experimental chemiluminescence records into wave-attached coordinates for reduced-order modeling \citep{mendibleDatadrivenModelingRotating2021}.
Based on our understanding, UnTWIST fits interpretable speed models offline from a full space--time snapshot matrix and takes the number of waves as an algorithmic input.
Those properties make it poorly suited to closed-loop mode-transition control, where the estimate must be causal and the wave count may change.
We therefore use a lightweight online tracker that discovers fronts from the field without a prescribed wave number and returns a single average speed for a shared frame.

At each environment step, we estimate $D$ from the just-completed trajectory segment.
We detect detonation fronts at each sub-step using the same gradient-thresholding method as in Section~\ref{sec:observation-space}.
We partition the segment into blocks of constant shock count, track each front across consecutive sub-steps by searching near its previous position (with periodic wrapping), and compute an instantaneous speed from the position difference.
We then average the speeds across all detected fronts within each usable block, and average again over blocks. Blocks with no shocks or only a single time sample are skipped.
If no usable estimate is obtained, we retain the previous velocity.
Near nucleation or merging of fronts the estimate can exhibit residual jitter. This jitter acts as observation noise and does not seem to affect the trained policies.
We then update the moving reference frame position $\Psi(t)$ during the next environment step according to the equation

\begin{equation}
    \Psi(t) = \bigl(\Psi(t_{\mathrm{prev}}) + D\,(t - t_{\mathrm{prev}})\bigr) \bmod L,
    \label{eq:moving_frame}
\end{equation}

\noindent where $t_{\mathrm{prev}}$ is the end time of the previous environment step and $D$ is the average speed over the previous environment step.
At $t = 0$, we initialize the reference frame velocity to $D = 1.71$ (the average mode-locked state speed, see \autoref{fig:wave-speed}) and subsequently update it each step using the tracked front speeds.
For certain visualizations, we use a slightly different moving reference frame, where we use a finer step size to update the position of the moving reference frame.
We label this finer-grained moving reference frame $\psi$.
Typically, we use a step size of $0.1$ time units to update $\psi$.

We compute observations in the moving reference frame by circularly shifting the state fields $u$ and $\lambda$ according to the position of $\Psi(t)$ (rounded to the nearest grid point).
After this shift, the observation vector is constructed as detailed in Section~\ref{sec:observation-space}.
We similarly shift control actions specified in the moving reference frame back to the stationary reference frame before being applied to the simulation: the injection pressure profile in the stationary reference frame is $u_p(x, t) = \tilde{u}_p\bigl((x - \Psi(t)) \bmod L\bigr)$, where $\tilde{u}_p$ is the control profile as seen by the agent.
Combined with the segmented injection pressure strategy, this enables the injection pressure segments to move along with the detonation waves, without necessitating a high actuation-update frequency.
The agent can thus adjust control on a slow timescale (the action period $\Delta t$), while the injection pattern automatically tracks the fast-timescale wave propagation through the geometric shift by $\Psi(t)$.

This timescale separation also shapes what experimental sensing would need to provide, if co-moving actuation is to be achieved in a laboratory setting.
Laboratory setups already resolve wave kinematics with high-speed chemiluminescence imaging \citep{kochModeLockedRotatingDetonation2020,kochDatadrivenSurrogatesRotating2021,mendibleDatadrivenModelingRotating2021} and with high-frequency wall-pressure or ionization measurements \citep{dengFeasibilityModeControl2018,bohonDynamicModeDecomposition2021}, at raw sampling rates that exceed the observation cadence used here.
Recent machine-learning diagnostics push toward online wave-mode and speed estimation, for example object-detection tracking of waves in down-axis images \citep{johnsonWaveDetectionTracking2023} and convolutional classification combined with ion-probe analysis at a few hertz \citep{johnsonUseConvolutionalNeural2023}.
The remaining sensing gap therefore lies in online processing rather than in raw measurement bandwidth.
The more severe barrier is actuation.
The policy and the estimate of $D$ are refreshed only every $\Delta t$, but the segmented injection pattern is shifted continuously according to~\eqref{eq:moving_frame}.
In the simulation, this shift is applied at every right-hand-side evaluation, so the laboratory-frame injection field is rewritten effectively continuously, independent of $\Delta t$.
In practice, a finite number of injectors limits this to a finite update rate: one update per grid-cell passage ($N = 512$) corresponds to roughly $3\,\mathrm{MHz}$ under the laboratory scaling above, which we regard as an upper bound (\autoref{tab:physical-rates}).
Even the far lower rates required by simple wave-attached templates (\autoref{fig:sss-ssm-3to4}) are well beyond valve-based actuation.
\subsection{Agent configurations}
\label{sec:agent-configs}

We study five controller configurations, one uniform and four segmented, listed in \autoref{tab:naming-conventions}.
The first letter denotes single-agent (S) or multi-agent (M) training, and the second denotes uniform (U) or segmented (S) injection pressure.
For segmented controllers, a third letter denotes a stationary (S) or moving (M) reference frame.
The uniform controller is denoted SU without a frame suffix because the stationary vs. moving reference frame distinction is only relevant for the sectioned state observations used by the segmented controllers.

For segmented control, a single agent outputs all $M=16$ injection-pressure values jointly.
The multi-agent formulation instead divides the same physical environment into 16 pseudo-environments with a shared policy, one scalar action per segment, and a shared global reward.
This parameter-sharing construction exploits translational invariance \citep{belusExploitingLocalityTranslational2019,suarezActiveFlowControl2025}.

\subsection{Baseline controllers}
Here we present the baseline controllers that we use for comparison
with SU. Both baseline controllers use uniform injection pressure and therefore have the same spatial actuation authority as SU.
This benchmark covers uniform actuation only. Independent azimuthal modulation gives the segmented controllers additional control authority, so they are not compared against these baselines.
We are not aware of an established RDE controller that senses the wave state and commands azimuthally resolved injection, which is what a baseline matched to the segmented configurations would require. Active RDE control demonstrated to date acts on a single global input, such as the propellant supply or the initiation process, and is typically scheduled rather than driven by the measured wave state \citep{dengFeasibilityModeControl2018,shengActiveDirectionControl2022,rongInvestigationCounterrotatingShock2022}.

\subsubsection{Two-step controllers}
Based on manual exploration through an interactive interface with the RDE simulation, we devised a general two-step controller strategy.
Starting from a mode-locked state, the controller rapidly decreases the injection pressure to a level below the reference pressure for the target state. The reference pressure is an injection pressure value that, when applied uniformly in the domain, reliably produces a mode-locked state with the target number of detonation waves. We find the reference pressure values by trial and error.
The controller then waits for a period of time and then increases the
injection pressure rapidly to the reference pressure for the target state.
The intention behind this controller is to rapidly destabilize the
system, and then re-stabilize it to the target state.

\begin{figure}[tbp]
    \centering
    \includegraphics[width=16cm, height=9cm]{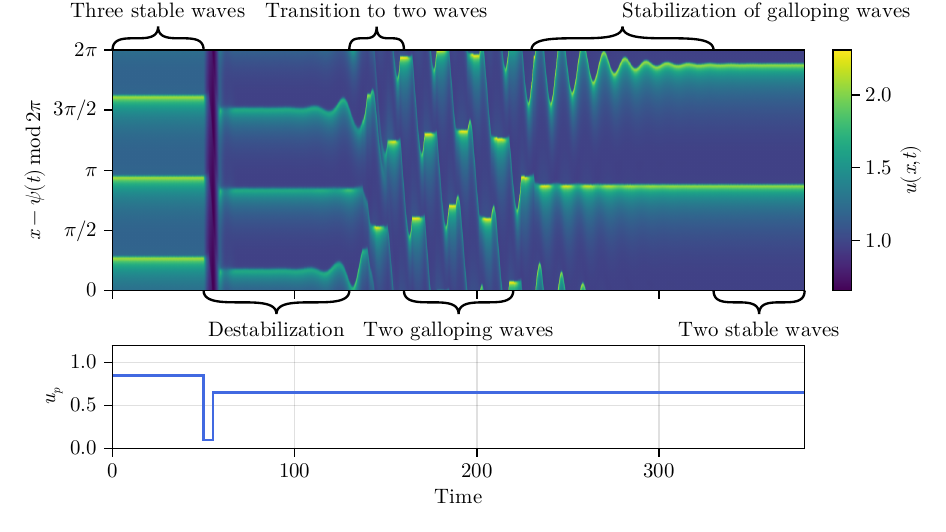}
\caption{Example simulation using the best two-step controller for targeting two detonation waves. The data in the top plot is shown with a moving reference frame with position $\psi$, as detailed in Section~\ref{sec:moving-frame}. After an initial drop and subsequent increase in injection pressure, the system eventually bifurcates from the three-detonation-wave mode-locked state into a galloping two-detonation-wave state, which ultimately settles into a mode-locked state with two waves. For a visual comparison of the stationary and moving reference frame, we refer to \autoref{fig:sample-run}.}
    \label{fig:two-step-example}
\end{figure}

For all four targeted mode-locked states,
we performed a grid search to find the best parameters for this two-step strategy.
An example of a simulation using a two-step controller is shown in \autoref{fig:two-step-example}.

\subsubsection{Simple PID controller}
In addition to the two-step controller, we implement a simple PID
controller.
The setpoint for the PID controllers is the reference pressure for
the target state.
The process variable is the injection pressure itself, not a measurement
of the detonation structure, so these controllers regulate pressure toward
the setpoint without feedback from the wave dynamics.
We constrain (clamp) the controller output to the range $[-1,1]$
and then pass it through the same linear action-control mapping as 
the output of the neural networks in the DRL controllers.

We performed a grid search over PID coefficients under the same
100-time-unit initialization hold and 0.5-time-unit resolution used at
evaluation, resulting in four tuned controllers, one for each
mode-locked state.

\subsection{Experimental setup}
\label{sec:experimental-setup}
\subsubsection{Training setups}

We train a separate policy for each target wave count $n_{\mathrm{target}}\in\{1,2,3,4\}$.
Training episodes last 500 time units, approximately 136 wave revolutions at the mean speed $D=1.71$.
At reset, the environment samples a mode-locked state and applies a random circular shift.
Both setups use one physical environment per training process, separate actor and critic networks with two hidden layers of 128 units, and rollout batches containing 2048 actions.
For MSS and MSM, the 2048 actions are distributed over 16 pseudo-environments, and the reported training-step count is the total number of pseudo-environment steps.

The uniform setup trains SU using the 16-entry Fourier-magnitude observation and one scalar action at the fixed period $\Delta t=10$.
We train six random seeds per target for 200,000 steps using $T_h=100$ and $\gamma=0.9$.
This dedicated setup uses an entropy coefficient of 0.01, four PPO epochs per rollout, and KL-divergence early stopping at 0.03.

The segmented setup trains SSS, SSM, MSS, and MSM with 16 control segments and the 66-entry sectioned observation.
For each method and target, we train five runs at each $\Delta t\in\{0.25,0.5,1,2,4,8\}$ for 800,000 steps.
These runs use $T_h=10$, $\gamma=1-\Delta t/T_h$, zero entropy coefficient, ten PPO epochs, and no KL-divergence early stopping.
The two setups share the remaining PPO settings summarized in \autoref{tab:setup-summary}.

We normalize observations element-wise using a running mean and standard deviation, and scale rewards by the running standard deviation of the discounted returns, following \citet{raffinStableBaselines3ReliableReinforcement2021}. Values are clipped to $[-10, 10]$, which never activates because all observations and rewards remain well within $\pm 10$.
Training was performed on CPUs (Intel Xeon E5 and Xeon Gold, 48--72 cores per node) on four machines, with up to 40 jobs run concurrently.
The mean wall-clock training time for the segmented runs was between six and seven hours.

\begin{table}[tbp]
    \centering
    \caption{Summary of the common and setup-specific simulation, environment, and PPO parameters.
    The segmented column applies to SSS, SSM, MSS, and MSM.}
    \renewcommand{\arraystretch}{1.08}
    \small
    \begin{tabular}{p{6.0cm}ll}
      \toprule
      \multicolumn{3}{l}{\textit{Common simulation and PPO settings}} \\
      \midrule
      Domain length $L$ & \multicolumn{2}{l}{$2\pi$} \\
      Grid cells $N$ & \multicolumn{2}{l}{$512$} \\
      Spatial scheme & \multicolumn{2}{l}{FV + MUSCL/MC + Rusanov} \\
      Time integration & \multicolumn{2}{l}{SSPRK33} \\
      CFL safety factor & \multicolumn{2}{l}{$0.62$} \\
      Action range & \multicolumn{2}{l}{$[-1,1]\to u_p\in[0,1.2]$} \\
      Smoothing time $\tau_{\mathrm{smooth}}$ & \multicolumn{2}{l}{$0.1$ time units} \\
      Episode duration & \multicolumn{2}{l}{500 time units} \\
      Actor architecture & \multicolumn{2}{l}{two 128-unit hidden layers, $\tanh$ activation} \\
      Critic architecture & \multicolumn{2}{l}{two 128-unit hidden layers, $\tanh$ activation} \\
      Weight initialization & \multicolumn{2}{l}{Orthogonal} \\
      Optimizer & \multicolumn{2}{l}{Adam} \\
      Learning rate & \multicolumn{2}{l}{$3\times10^{-4}$} \\
      GAE $\lambda$ & \multicolumn{2}{l}{$0.95$} \\
      PPO clip range & \multicolumn{2}{l}{$0.2$} \\
      Value-function coefficient & \multicolumn{2}{l}{$0.5$} \\
      Gradient-norm limit & \multicolumn{2}{l}{$0.5$} \\
      Minibatch size & \multicolumn{2}{l}{$64$} \\
      Rollout batch size & \multicolumn{2}{l}{$2048$ actions} \\
      Parallel physical environments & \multicolumn{2}{l}{$1$} \\
      Advantage normalization & \multicolumn{2}{l}{Yes} \\
      Observation normalization & \multicolumn{2}{l}{Yes} \\
      Reward normalization & \multicolumn{2}{l}{Yes} \\
      \midrule
      \textit{Setup-specific settings} & \textit{Uniform} & \textit{Segmented} \\
      \midrule
      Observation & 16 Fourier magnitudes & 66 sectioned entries \\
      Control segments & $1$ & $16$ \\
      Pseudo-environments & $1$ & $1$ (SSS/SSM), $16$ (MSS/MSM) \\
      Reference frame & N/A & stationary or moving \\
      Action period $\Delta t$ & $10$ & $\{0.25,0.5,1,2,4,8\}$ \\
      Physical reward horizon $T_h$ & $100$ & $10$ \\
      Discount rate $\gamma$ & $1-\Delta t/T_h=0.9$ & $1-\Delta t/T_h$ \\
      Total training steps & $200{,}000$ & $800{,}000$ \\
      Seeds/runs per target and setting & $6$ & $5$ \\
      Entropy coefficient & $0.01$ & $0$ \\
      PPO epochs per rollout & $4$ & $10$ \\
      KL-divergence early stopping & $0.03$ & No \\
      \bottomrule
    \end{tabular}
    \label{tab:setup-summary}
\end{table}

\subsubsection{Transition evaluation}
\label{sec:transition-evaluation}

We evaluate transitions from each of the three off-target mode-locked states for every target wave count.
Each initial state is a precomputed converged solution of \eqref{eq:rde_equations}, circularly shifted at random for each of five trials.
The simulation lasts 600 time units.
The injection pressure associated with the initial mode is held fixed for the first 100 time units. The tested controller is then activated.

Evaluation uses the stability component of the training reward with $\alpha_v=1$.
A transition succeeds when the target wave count is present and the stability reward remains above 0.9 for a continuous interval of 20 time units.
We define transition time as the interval from controller activation to the beginning of the qualifying stability interval.
A failed trial has no finite transition time and is excluded from conditional mean transition times. Success rate must therefore be reported alongside transition time.
We characterize each initial-to-target pair by these two metrics throughout: success rate measures reliability across trials, while conditional transition time measures speed as the mean over successful trials.

For the matched uniform benchmark, SU, two-step, and PID controllers are assessed with a 0.5-time-unit resolution. The SU action is recomputed only at its trained period $\Delta t=10$ and held between updates. All three controllers share the same initial states, activation time, horizon, and success criterion.
For each target, we select one of the six SU seeds by highest mean success over the three initial states and then lowest mean transition time among successful trials.
The two-step and PID grid searches likewise provide one controller per target, shared across the three initial states.

Evaluation of the segmented controllers retains each policy's trained action period.
For the transition-time comparison in \autoref{fig:segmented-performance}(a), we select one trained model per method and target by highest mean success over the three initial states and then lowest conditional mean transition time.
All other segmented figures aggregate every trained run.

\section{Results and Discussion}
\label{sec:results}

\subsection{Matched uniform-control benchmark}

SU succeeds in all five trials for all 12 pairs.
The two-step and PID controllers also succeed in all five trials except on $3\to4$, where two-step misses one trial and PID succeeds in only one of five.
This indicates that $3\to4$ is particularly difficult to control.
The difficulty agrees with the more chaotic picture in the three- and four-wave regions of \autoref{fig:wave-speed}: simply increasing $u_p$ from the three-wave mode-locked state leads to a transition to chaos, and the controller must learn either to escape that regime or to avoid it entirely in order to reach the four-wave mode-locked state.

\begin{figure}[tbp]
    \centering
    \includegraphics[width=16cm]{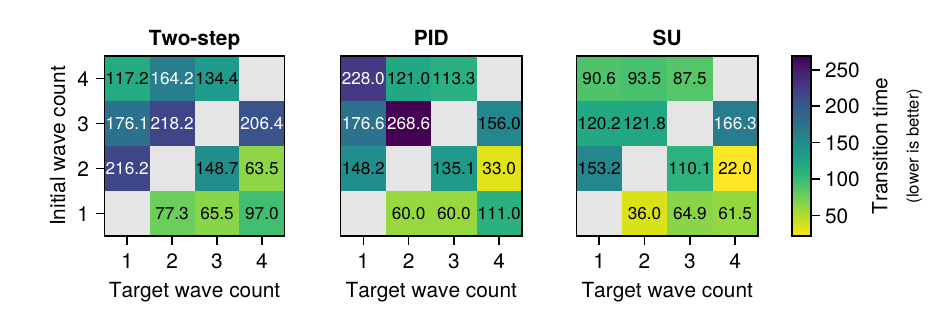}
    \caption{Matched uniform-control benchmark: conditional mean transition time over successful trials (five trials per pair).
    Rows indicate the initial wave count and columns indicate the target wave count. The diagonal entries are unused.
    All three controllers are assessed with a 0.5-time-unit resolution after a 100-time-unit uncontrolled initialization.
    SU actions are updated every 10 time units and held between updates.
    The displayed SU policy is selected separately for each target from six seeds by highest mean success and then lowest conditional mean transition time.
    SU has a 100\% success rate for all transitions, whereas the two-step and PID baselines struggle on $3\to4$, where they reach only 80\% and 20\% success, respectively.
    The medians over the initial-to-target pairs with finite transition times are 92.05 for SU, 141.55 for two-step, and 128.05 for PID.}
    \label{fig:uniform-baselines}
\end{figure}

The conditional mean transition times in \autoref{fig:uniform-baselines} use the common evaluation protocol described in Section~\ref{sec:transition-evaluation}.
Ranking each pair first by success rate and then by conditional transition time gives 12 wins, no losses, and no ties for SU against the two-step controller, and 10 wins and 2 losses against PID.
SU is faster in all 12 pairs jointly successful with the two-step controller.
PID is faster by a small margin on $2\to1$ (148.2 versus 153.2) and $1\to3$ (60.0 versus 64.9).
These two margins are small, and the comparison is not matched in decision rate: SU updates only every $\Delta t=10$ and holds the action between updates, whereas the baselines may adjust at the $0.5$-time-unit evaluation resolution.
We adopted $\Delta t=10$ for the SU setup after preliminary runs at shorter periods were less reliable under the present reward and PPO settings.
DRL methods that better handle delayed, multiscale credit assignment might support a shorter action period and narrow these gaps.
On $3\to4$, SU's mean time is 166.3, compared with 206.4 for the four successful two-step trials. PID is unreliable on this pair, but its single successful trial is slightly faster at 156.0.
The medians over the initial-to-target pairs with finite transition times are 92.05 for SU, 141.55 for two-step, and 128.05 for PID.

\begin{figure}[tbp]
    \centering
    \includegraphics[width=16cm, height=9cm]{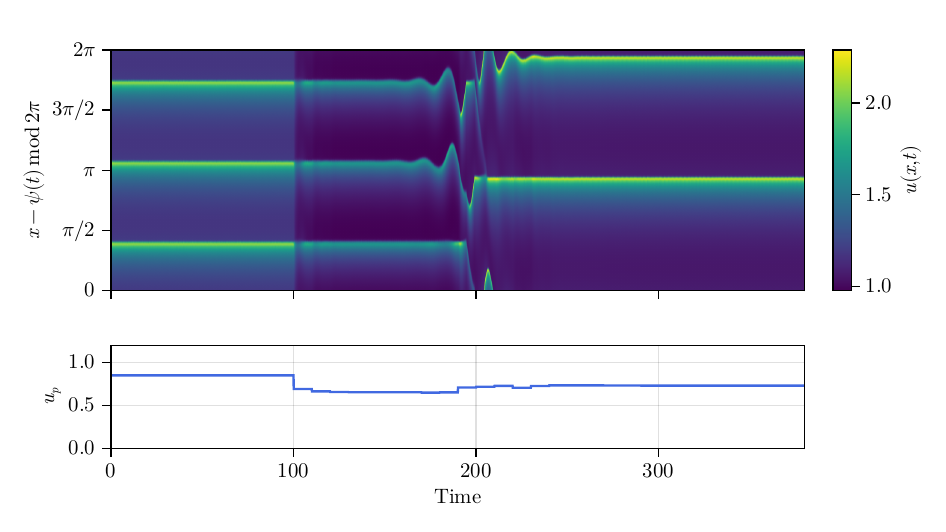}
    \caption{Example simulation using the selected SU controller for targeting two detonation waves, starting from the same three-wave initial state as \autoref{fig:two-step-example}. The data in the top plot is shown with a moving reference frame with position $\psi$, as detailed in Section~\ref{sec:moving-frame}. The controller is activated at $t=100$ and reaches the two-wave mode-locked state after a transition time of $121.5$ time units, compared with $218.2$ for the two-step controller mean in \autoref{fig:uniform-baselines}.}
    \label{fig:uniform-3to2}
\end{figure}

A representative SU control trajectory for the $3\to2$ transition is shown in \autoref{fig:uniform-3to2}, for direct comparison with the two-step trajectory in \autoref{fig:two-step-example}.

Uniform actuation can influence only the global injection pressure, so many transitions require a long sequence before the reward reflects the effect of an action.
The longer SU reward horizon was chosen to cover this slower response.
Despite this hypothesized temporal-credit challenge, the selected SU policies remain reliable and faster than the two baselines in the matched evaluation.

\subsection{Segmented-controller performance}

\begin{figure}[tbp]
    \centering
    \includegraphics[width=16cm]{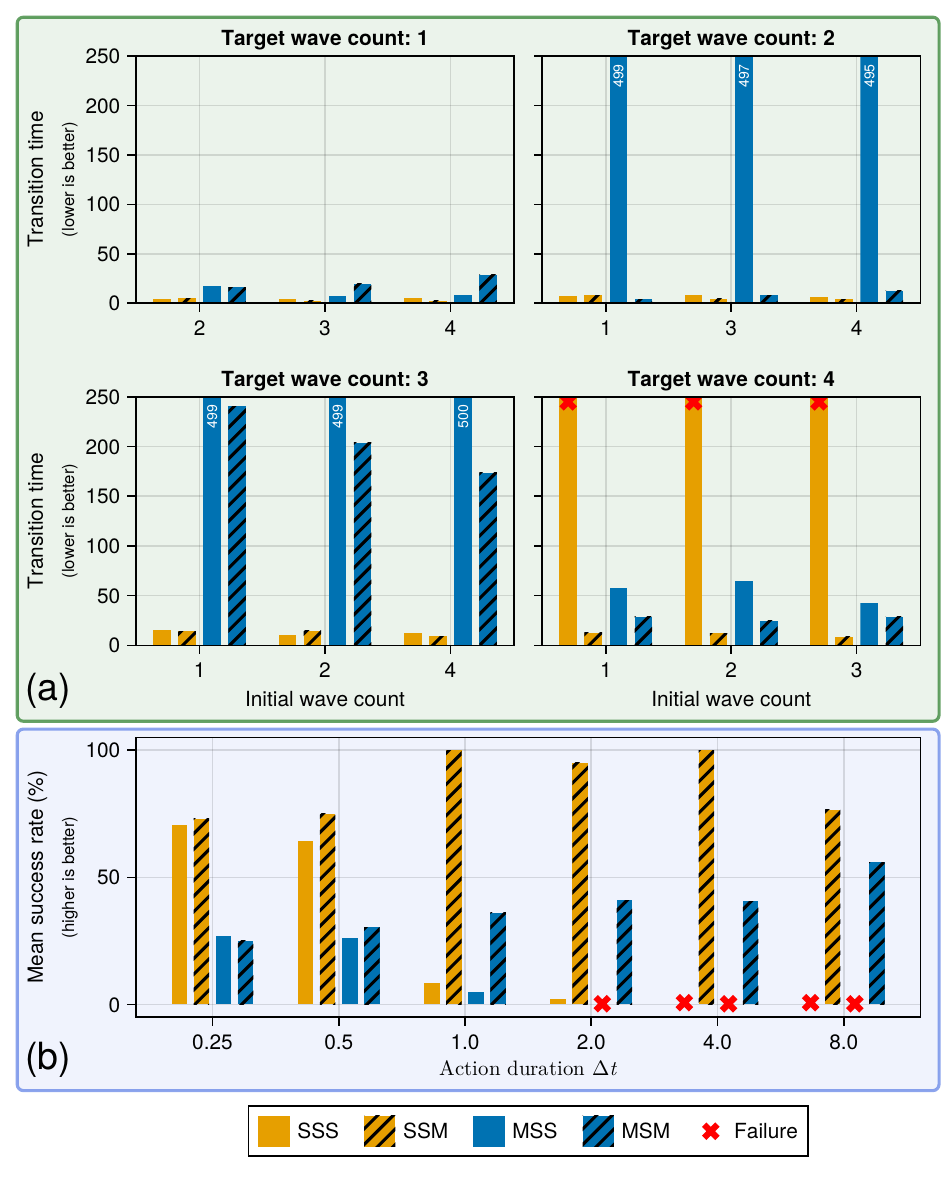}
    \caption{Performance of segmented DRL controllers.
    (a)~Conditional mean transition times for the best model selected separately for each method and target by highest mean success and then lowest conditional mean time.
    Each group corresponds to one target and contains the three off-target initial wave counts.
    A full-height bar with a red cross denotes no successful trial, not a transition measured at the plotted upper limit.
    (b)~Mean success rate versus action period, averaged over five runs, four targets, and the three off-target initial states.
    Each policy is evaluated at its trained action period using five trials per initial-to-target pair.
    Lower transition time and higher success rate are preferable.}
    \label{fig:segmented-performance}
\end{figure}

The segmented-controller study is summarized in \autoref{fig:segmented-performance}.
The selected segmented policies reach their targets in substantially less time than the uniform SU controllers in \autoref{fig:uniform-baselines}, showing that the learned policies utilize the additional spatial authority of azimuthally resolved injection.
Panel~(a) shows the best model selected for each method and target.
The selected SSM and MSM models complete all 12 transition pairs, but SSM is markedly faster for transitions to three waves.
The selected SSS models complete all transitions to one, two, and three waves but fail for every transition to four waves.
The selected MSS models are less reliable for targets two and three and have conditional transition times near the 600-time-unit evaluation horizon in several of those pairs.
The four-wave target is especially difficult for SSS. \autoref{fig:wave-speed} indicates a narrow stable high-$u_p$ range near regions of hysteresis and non-mode-locked or chaotic dynamics, which is consistent with SSS failing on every transition to four waves.

Panel~(b) includes all trained runs and exposes sensitivity to the action period.
SSM has mean success rates from 73\% to 100\% across all six action periods, reaching 100\% at $\Delta t=1$ and $4$.
SSS performs well only at the two shortest periods (70.7\% and 64.3\%) and falls to 8.3\% at $\Delta t=1$ before approaching zero.
MSM improves from 25.0\% at $\Delta t=0.25$ to 56.0\% at $\Delta t=8$, whereas MSS reaches 27.0\% at best and has no successes for $\Delta t\geq2$.
The moving-frame formulations therefore remain effective over a broader range of action periods than their stationary counterparts.

\subsection{Reference-frame and agent effects}

\begin{figure}[tbp]
    \centering
    \includegraphics[width=14cm]{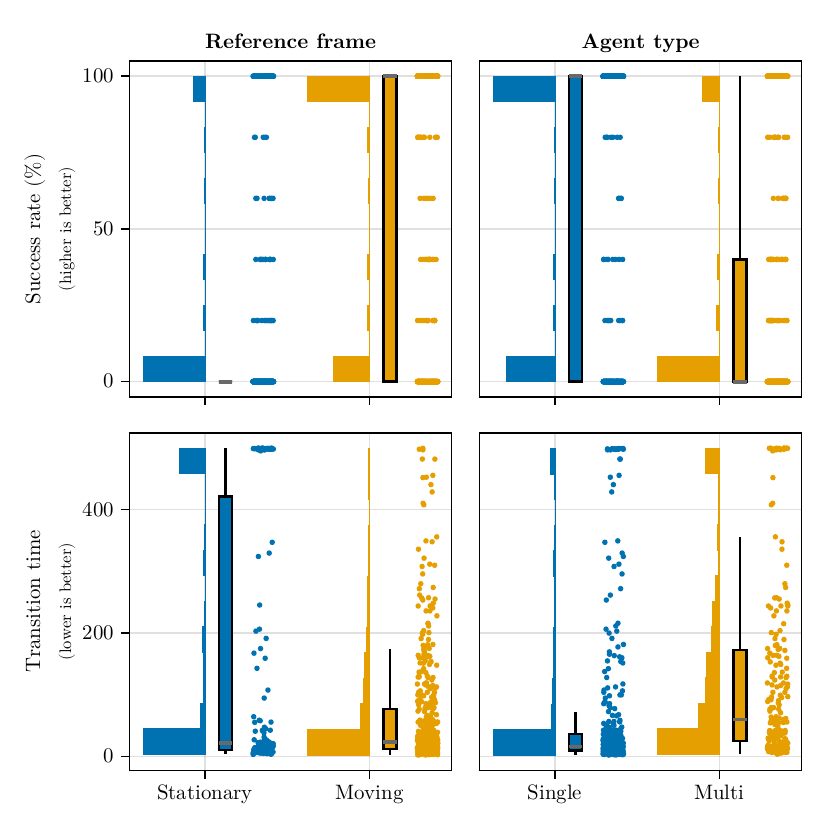}
    \caption{Marginal distributions for the two segmented-controller design factors: reference frame (left) and agent structure (right).
    The top row shows success rate and includes all trained-model and initial-to-target pairs.
    The bottom row shows transition time and includes successful pairs only.
    Each raincloud combines a histogram, boxplot, and jittered observations.
    Success rates are discrete percentages from five trials ($0$, $20$, \ldots, $100\%$).
    Each panel pools over the other design factor and over targets, seeds, and action periods.}
    \label{fig:config-facet-comparison}
\end{figure}

The segmented-controller evaluations are pooled by reference frame and agent structure in \autoref{fig:config-facet-comparison}.
The success distributions include every trained-model and initial-to-target pair, including failures, while the transition-time distributions include successful pairs only.
The main effects are clear. Controllers in the moving frame are more reliable than their stationary counterparts, and single-agent training is more reliable than multi-agent parameter sharing.
The moving-frame means are 86.6\% for SSM and 38.2\% for MSM, against 24.2\% for SSS and 9.7\% for MSS.
The stationary distributions are strongly zero-inflated: 72.5\% of SSS pairs and 86.7\% of MSS pairs have zero success, giving both methods a median success of zero despite successful subsets.

\subsection{Resolved robustness and training}

\begin{figure}[tbp]
    \centering
    \includegraphics[width=16cm]{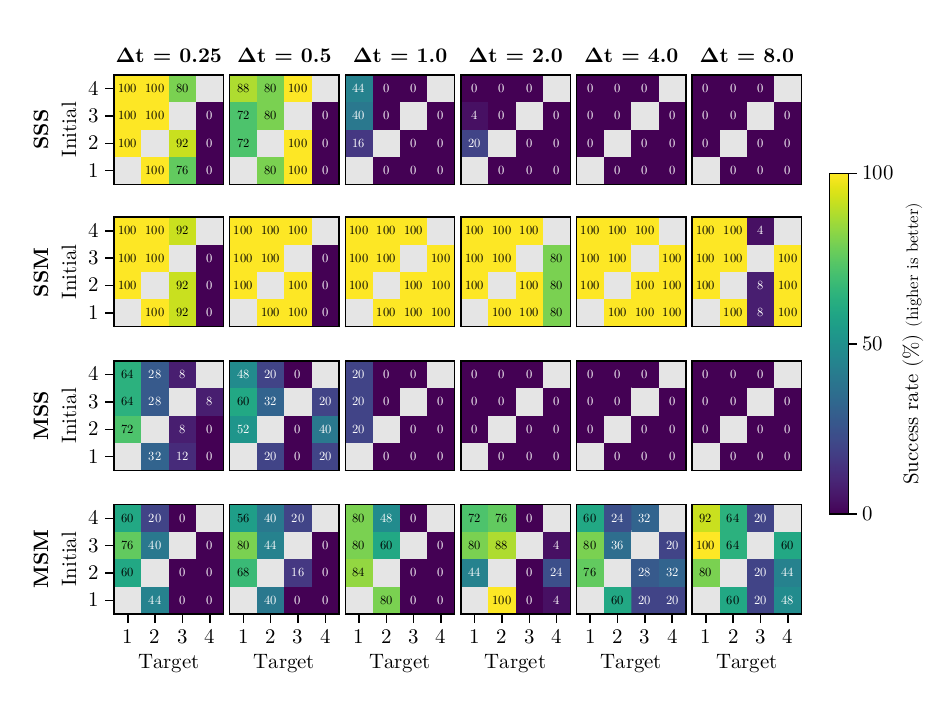}
    \caption{Segmented-controller success rates resolved by method (figure rows), action period $\Delta t$ (figure columns), initial wave count (rows within each panel), and target wave count (columns within each panel).
    Each entry is the mean percentage over five trained runs, with five evaluation trials per run and initial-to-target pair.
    The diagonal entries are unused because the initial and target modes coincide.}
    \label{fig:success-rate-matrices}
\end{figure}

Success rates are resolved by initial state, target, and action period in \autoref{fig:success-rate-matrices}.
SSM has the broadest coverage, including complete mean success at $\Delta t=1$ and $4$.
Its remaining weaknesses are concentrated in transitions to four waves at the shortest periods and to three waves at $\Delta t=8$.
SSS succeeds mainly at short action periods and does not control the four-wave target reliably.
MSS fails broadly, while MSM improves on MSS and retains partial success at $\Delta t=8$.
The corresponding conditional transition-time matrices are provided in Appendix~\ref{app:transition-time-matrices}. The blank entries have no successful transition.

\begin{figure}[tbp]
    \centering
    \includegraphics[width=16cm]{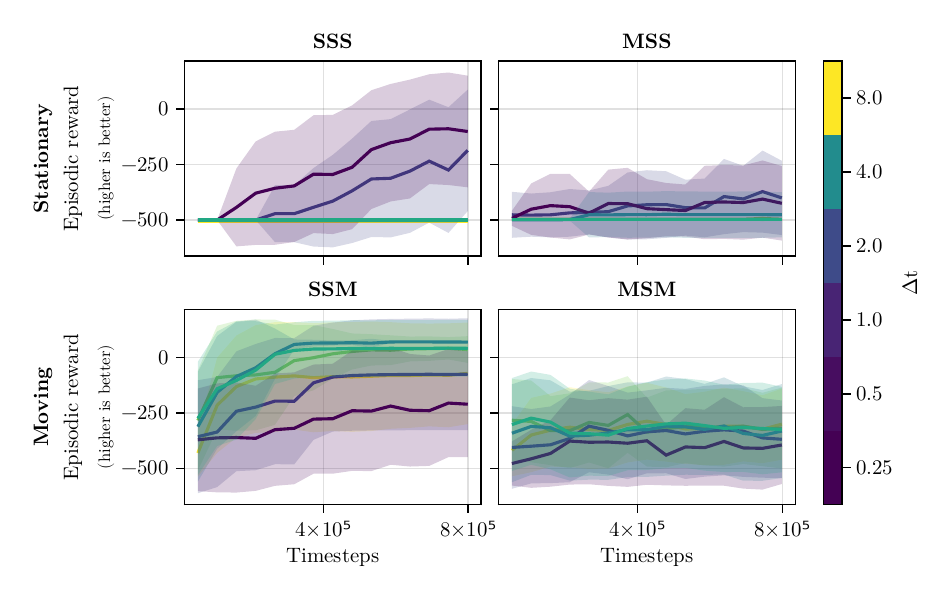}
    \caption{Evaluation episode reward during training for the four segmented controllers.
    The top row contains stationary-frame SSS and MSS, and the bottom row contains moving-frame SSM and MSM.
    Curves show the mean over five runs and four target wave counts. The shaded bands show one standard deviation, and color identifies the action period $\Delta t$.
    For MSS and MSM, the horizontal axis counts total pseudo-environment steps.}
    \label{fig:training-curves-reward}
\end{figure}

The corresponding evaluation-reward trajectories during training are shown in \autoref{fig:training-curves-reward}.
SSM learns for most action periods and typically plateaus near $4\times10^5$ pseudo-environment steps.
SSS learns primarily at short action periods, MSS remains nearly flat, and MSM improves on MSS with stronger curves at longer action periods.
The SSM curve for $\Delta t=0.25$ is still increasing at the end of training and finishes below the corresponding SSS curve.
SU training curves are not shown because the SU setup uses different observations, horizon, action period, training duration, and PPO settings (Section~\ref{sec:experimental-setup}).

\subsection{Why the moving frame helps}

At fixed physical reward horizon $T_h$, reducing $\Delta t$ increases the decision-sequence length $T_h/\Delta t$.
More neighboring actions can then contribute to each delayed outcome, which we expect to make temporal credit assignment more difficult \citep{karimiDynamicDecisionFrequency2023,metelliControlFrequencyAdaptation2020,pignatelliSurveyTemporalCredit2023}.
This challenge is particularly relevant in the RDE environment, where mode transition is slow while the controlled detonation fronts translate rapidly.

In stationary coordinates, a segmented controller must update quickly enough to recreate a wave-relative action as the fronts pass fixed segments.
The moving transformation instead presents a quasi-steady wave pattern to the policy and transports the selected segmented action continuously between control updates.
The policy therefore selects the slowly varying spatial modulation while the coordinate transformation handles fast translation.
This agent--environment reformulation reduces the effective depth of the tracking problem without shortening the physical reward horizon.
It resembles temporal abstraction in its effect on credit assignment \citep{suttonMDPsSemiMDPsFramework1999}, although it is implemented as a symmetry transformation rather than a hierarchy of learned actions.

  Segmented actuation may also create a compounding advantage. Greater spatial authority can produce target-relevant local changes sooner, shortening the delay from an action to an informative reward and thereby making the control law easier to learn, which in turn can shorten transitions further. This mechanism is one possible contributing explanation for the drastically lower transition times of the segmented controllers.

\subsection{Learned control mechanisms}

\begin{figure}[tbp]
    \centering
    \includegraphics[width=16cm, height=20cm]{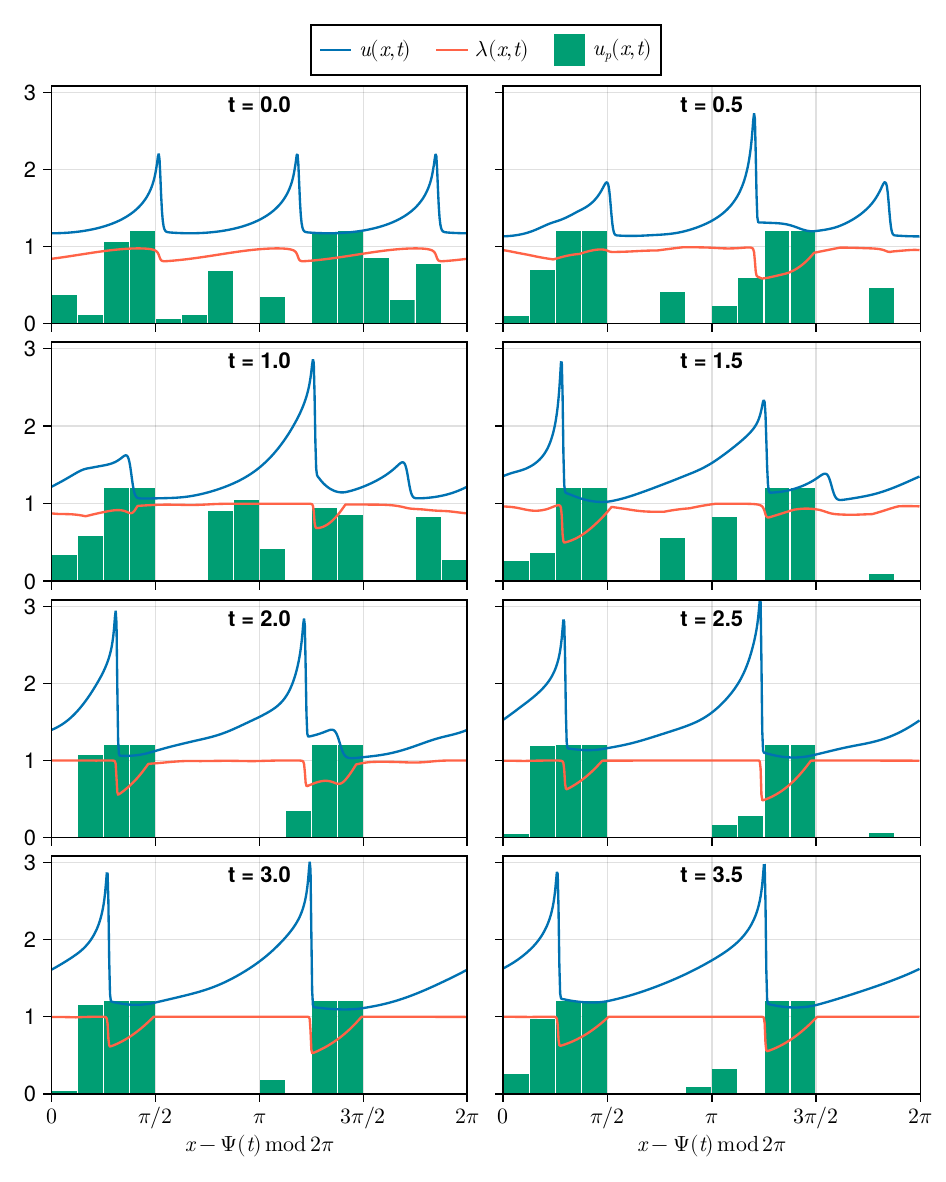}
    \caption{Internal energy $u$, combustion progress $\lambda$, and injection pressure $u_p$ during a selected SSM transition from three to two waves, shown in the moving reference frame defined by \eqref{eq:moving_frame}.
    The policy applies an asymmetric pressure distribution that weakens one wave and guides the system toward the target two-wave mode.}
    \label{fig:control-snapshots}
\end{figure}

A selected SSM transition from three to two waves is shown in \autoref{fig:control-snapshots}.
The controller applies low injection pressure near the middle detonation wave and higher pressure near the other two.
The middle wave weakens and is absorbed by its left neighbor, after which the controller maintains a stable two-wave pattern.

\begin{figure}[tbp]
    \centering
    \includegraphics[width=16cm]{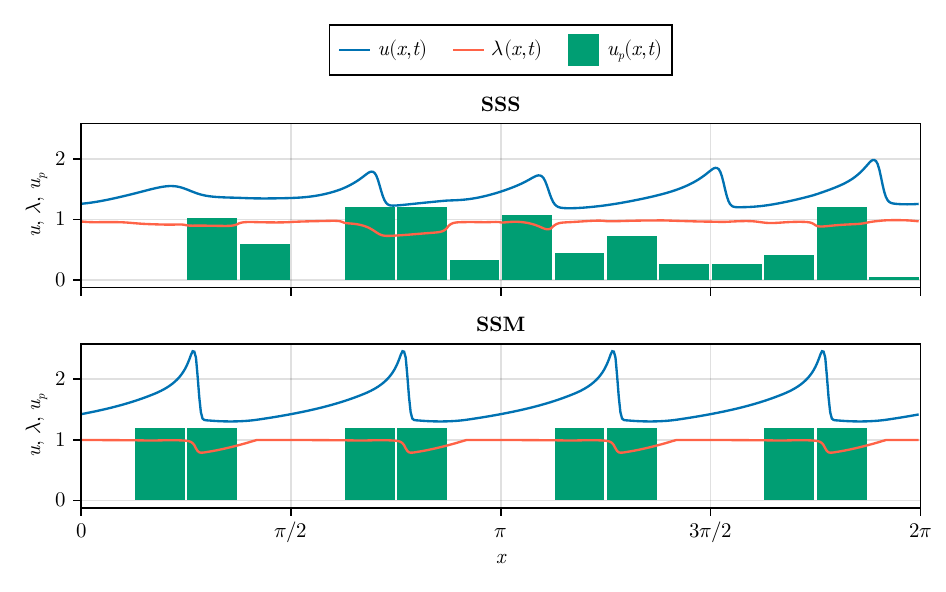}
    \caption{Internal energy $u$, combustion progress $\lambda$, and injection pressure $u_p$ at $t=105$, 85 time units after controller activation, for two selected segmented controllers targeting the $3\to4$ transition.
    SSS with $\Delta t=0.25$ is shown above, and SSM with $\Delta t=4$ is shown below. Both rows use the laboratory-frame coordinates.
    SSS remains irregular at this instant, whereas SSM has reached a four-wave state with the waves near the centers of four high-$u_p$ regions.}
    \label{fig:sss-ssm-3to4}
\end{figure}

Stationary and moving segmented control for the $3\to4$ transition are contrasted in \autoref{fig:sss-ssm-3to4}.
The selected SSM controller establishes an approximately four-lobed, bang--bang injection template in its moving coordinates before the fourth wave has fully formed.
The transformation transports this slowly varying template around the laboratory domain.
An additional wave forms, and the four waves align near the high-$u_p$ regions.
The selected SSS trajectory remains irregular, with its wave count fluctuating among three, four, and five.
Because SSS segments are fixed, its policy must recreate the wave-relative injection pattern through rapid updates as the fronts travel across the segments.
In SSM, the coordinate transformation carries that transport, so the policy only selects the slowly varying template.

Open-loop controllers are often easier to implement than closed-loop ones, and they are closer to the scheduled actuation used in the RDE control demonstrated to date \citep{dengFeasibilityModeControl2018,shengActiveDirectionControl2022,rongInvestigationCounterrotatingShock2022}.
The low temporal complexity of the SSM action suggests that the policy could be distilled into an open-loop deterministic controller that prescribes and transports a similar four-lobed template.
Because the template is approximately bang--bang, a fixed injector would switch about eight times per revolution, or roughly $45\,\mathrm{kHz}$ under the laboratory scaling in \autoref{tab:physical-rates}, far below the grid-resolution bound but still to our knowledge not realizable with the current RDE injection hardware.

\subsection{Single-agent and multi-agent control}

The multi-agent decomposition is designed for high-dimensional translationally invariant systems with local actuation effects and informative local rewards \citep{belusExploitingLocalityTranslational2019}.
The RDE objective is instead global: every pseudo-agent receives the same wave-count and stability reward, and there is no obvious segment-level success signal.
Actions must also be coordinated across segments to decide which fronts to weaken or reinforce.
The single-agent segmented controller observes the full field summary and produces all 16 actions jointly, allowing coordination through shared hidden features.

These differences provide a plausible explanation for the lower MSS and MSM reliability, but the experiments do not isolate reward locality from coordination.
Adding normalized segment indices and sinusoidal positional encodings did not improve the multi-agent results, suggesting that missing absolute position is not the principal limitation.
Circular convolutional encoders, structured action decoders, message passing, or recurrent policies could provide stronger spatial or temporal inductive biases.
Such architectures may improve representation and coordination, but they do not by themselves assign actuator-specific credit from the global reward.
In this reduced-order setting, the 16-dimensional single-agent action remains tractable and SSM is the most reliable segmented formulation.

\section{Conclusion}
\label{sec:conclusion}

We presented a first application of deep reinforcement learning to control mode transitions in a reduced-order one-dimensional RDE model.
Under matched uniform actuation and a common evaluation protocol, the selected SU policies succeeded in every evaluated transition.
Ranking each pair first by success rate and then by conditional transition time, SU is faster than the two-step controller in every pair and faster than PID in ten of twelve pairs.
PID is faster on two pairs, both by a small margin.
The main reliability gap is the $3\to4$ transition, which only SU completes in every trial.
This result establishes that DRL can provide effective uniform RDE control within the present model and protocol.

The segmented study provides the main methodological result.
Expressing observations and spatially distributed actions in a moving reference frame makes the detonation structure quasi-steady from the policy's perspective.
SSM consequently learns reliable transitions over a broader range of action periods than SSS, while the moving-frame multi-agent formulation also improves on its stationary counterpart.
The results support symmetry-aware scale separation as a way that may reduce temporal credit-assignment difficulty in multiscale flow control.

\paragraph{Future work.}
Although the present study benefits from the low computational cost of the one-dimensional model, scaling the approach to two- or three-dimensional simulations will require improved sample efficiency and better reuse of learned policies.
In that setting, rapidly updating segmented injection pressure may also be less realistic as a direct actuation mechanism, but it can still serve as a useful tool for probing RDE dynamics and for evaluating control concepts.
Future work should compare architectures with stronger inductive biases for the annular domain.
A circular one-dimensional convolutional encoder could treat $u$ and $\lambda$ as co-located channels and encode periodic spatial adjacency, while a structured convolutional decoder or message passing on a ring graph could encode relationships and coordination among segmented action outputs \citep{jeonInductiveBiaseddeepReinforcement2025,kurzInvariantControlStrategies2025}.
Stacked observations or recurrent networks could provide temporal context when the instantaneous observation is insufficient.
A systematic comparison could establish how these architecture choices interact with the stationary and moving reference frames and with the single- and multi-agent configurations studied here.
Finally, it would be valuable to explore richer physics models \citep{kochModelingThermodynamicTrends2020a} and more engineering-relevant objectives, for example optimizing efficiency subject to thrust and stability constraints.

\appendix
\section{Segmented transition-time matrices}
\label{app:transition-time-matrices}

\begin{figure}[H]
    \centering
    \includegraphics[width=16cm]{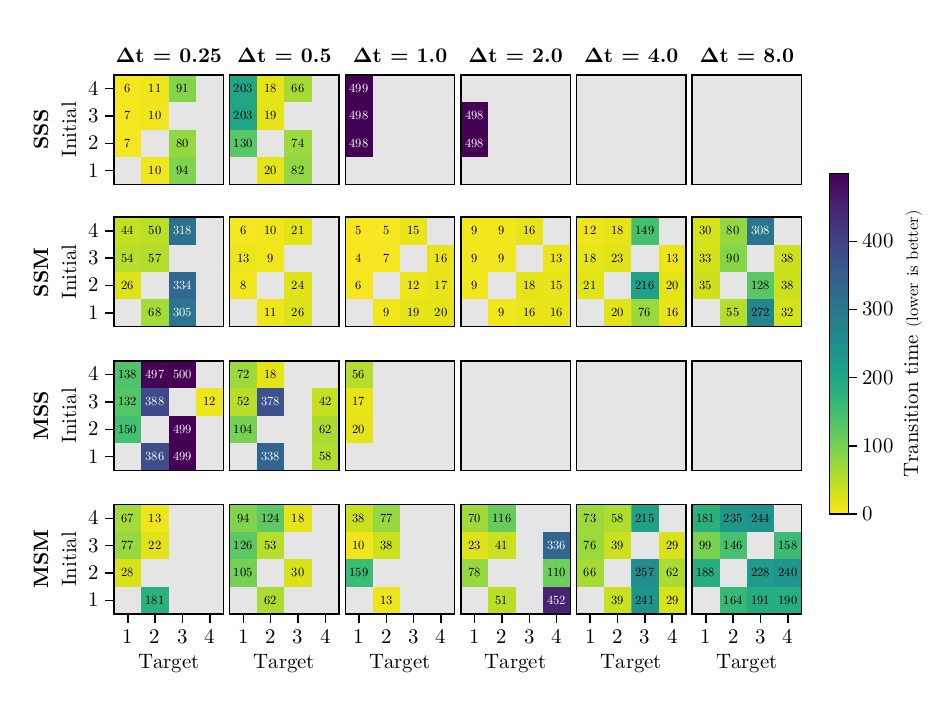}
    \caption{Conditional mean transition times for the four segmented controllers, resolved by method (figure rows), action period $\Delta t$ (figure columns), initial wave count (rows within each panel), and target wave count (columns within each panel).
    Each value averages the per-run conditional means over the trained runs with at least one successful trial in that cell.
    Blank entries have no successful transition, and diagonal entries are unused.}
    \label{fig:transition-time-matrices}
\end{figure}

The conditional transition-time counterpart to the success-rate matrices in \autoref{fig:success-rate-matrices} is provided in \autoref{fig:transition-time-matrices}.
These values characterize speed only where at least one evaluation succeeds and should therefore be interpreted together with the corresponding success rates.

\clearpage

\section*{Acknowledgements}
We used large language model (LLM)–based tools to assist with language editing and manuscript preparation.
The authors developed and verified all technical content, interpretations, and conclusions, and reviewed and validated all LLM-suggested edits.

\section*{Open source code, data and animations}
The code for the RDE simulator is available at \url{https://github.com/KristianHolme/RDE.jl}.
The code defining the DRL environment is available at \url{https://github.com/KristianHolme/RDE_Env.jl}.
The DRL implementation used in this work is available at \url{https://github.com/KristianHolme/Drill.jl}.
Data used to produce the figures in this paper are openly available~\citep{holmeDRL_RDE_data2026}.
We will make the code~\citep{kristianholmeKristianHolmeDRL_RDE_paper_code2026} for setting up and analyzing the training runs and producing all figures in this paper available at \url{https://github.com/KristianHolme/DRL_RDE_paper_code} upon publication. The code repository will contain information on how to download and use the openly available data.
We will provide technical support for the software through the respective GitHub issue tracking systems.
A series of videos illustrating the main dynamics of the RDE system with different controllers is available at \url{https://youtube.com/playlist?list=PLQEazQ69wnSRKEf-dCdlaa7-BnQkzhvGD&si=GQ_ZTh9pKbFSBbRW}.

\bibliographystyle{unsrtnat-doi.bst}
\bibliography{DRL-RDE-report}

\end{document}